\documentclass[journal]{IEEEtran} 
\IEEEoverridecommandlockouts
\usepackage{cite}
\usepackage{amsmath,amssymb,amsfonts}
\usepackage{algorithmic}
\usepackage{graphicx}
\usepackage{textcomp}
\usepackage{xcolor}
\def\BibTeX{{\rm B\kern-.05em{\sc i\kern-.025em b}\kern-.08em
    T\kern-.1667em\lower.7ex\hbox{E}\kern-.125emX}}
\usepackage[ruled,lined]{algorithm2e}
\usepackage[colorlinks]{hyperref}
\usepackage{multirow}
\usepackage{bbm}
\begin{document}

\title{DIAS: A Dataset and Benchmark for Intracranial Artery Segmentation in DSA sequences\\

\thanks{This work was supported in part by the National Natural Science Foundation of China under Grant 62376038 and in part by the National Key Research and Development Program of China under Grant 2018AAA0102600. }
}


\author{
    \IEEEauthorblockN{Wentao Liu\textsuperscript{1}}
    , \IEEEauthorblockN{Tong Tian\textsuperscript{2}}
    , \IEEEauthorblockN{Lemeng Wang\textsuperscript{1}}
    , \IEEEauthorblockN{Weijin Xu\textsuperscript{1}}
    , \IEEEauthorblockN{Lei Li\textsuperscript{3}}
    , \IEEEauthorblockN{Haoyuan Li\textsuperscript{1}}
    , \IEEEauthorblockN{Wenyi Zhao\textsuperscript{1}}
    , \IEEEauthorblockN{Siyu Tian\textsuperscript{4}}
    , \IEEEauthorblockN{Xipeng Pan\textsuperscript{5}}
   , \IEEEauthorblockN{Yiming Deng\textsuperscript{3}}
    , \IEEEauthorblockN{Feng Gao\textsuperscript{3,*}}
    , \IEEEauthorblockN{Huihua Yang\textsuperscript{1,5,*}}
    , \IEEEauthorblockN{Xin Wang\textsuperscript{6}}
    , \IEEEauthorblockN{Ruisheng Su\textsuperscript{7,8}}
    \\ \IEEEauthorblockA{\textsuperscript{1}\textit{School of Artificial Intelligence, Beijing University of Posts and Telecommunications, Beijing, China}}\\
     \IEEEauthorblockA{\textsuperscript{3}\textit{State Key Laboratory of Structural Analysis, Optimization and CAE Software for Industrial Equipment, School of Aeronautics and Astronautics, Dalian University of Technology, Dalian, China}}\\
    \IEEEauthorblockA{\textsuperscript{2}\textit{Department of Interventional Neuroradiology, Beijing Tiantan Hospital, Capital Medical University, Beijing, China}}\\
    
\IEEEauthorblockA{\textsuperscript{4}\textit{Ultrasonic Department, The Fourth Hospital of Hebei Medical University and Hebei Tumor Hospital, Shijiazhuang, China}}\\

     \IEEEauthorblockA{\textsuperscript{5}\textit{School of Computer Science and Information Security, Guilin University of Electronic Technology, Guilin, China}}\\
    \IEEEauthorblockA{\textsuperscript{6}\textit{Department of Radiology, The Netherlands Cancer Institute, Amsterdam, The Netherlands}}\\
    \IEEEauthorblockA{\textsuperscript{7}\textit{Department of Radiology \& Nuclear Medicine, Erasmus MC, University Medical Center Rotterdam, The Netherlands}}\\
     \IEEEauthorblockA{\textsuperscript{8}\textit{Medical Image Analysis group, Department of Biomedical Engineering, Eindhoven University of Technology, Eindhoven, the Netherlands}}\\
    \IEEEauthorblockA{liuwentao@bupt.edu.cn}
}

\maketitle
\begin{abstract}
The automated segmentation of Intracranial Arteries (IA) in Digital Subtraction Angiography (DSA) plays a crucial role in the quantification of vascular morphology, significantly contributing to computer-assisted stroke research and clinical practice. Current research primarily focuses on the segmentation of single-frame DSA using proprietary datasets. However, these methods face challenges due to the inherent limitation of single-frame DSA, which only partially displays vascular contrast, thereby hindering accurate vascular structure representation. In this work, we introduce DIAS, a dataset specifically developed for IA segmentation in DSA sequences. We establish a comprehensive benchmark for evaluating DIAS, covering full, weak, and semi-supervised segmentation methods. Specifically, we propose the vessel sequence segmentation network, in which the sequence feature extraction module effectively captures spatiotemporal representations of intravascular contrast, achieving intracranial artery segmentation in 2D+Time DSA sequences. For weakly-supervised IA segmentation, we propose a novel scribble learning-based image segmentation framework, which, under the guidance of scribble labels, employs cross pseudo-supervision and consistency regularization to improve the performance of the segmentation network. Furthermore, we introduce the random patch-based self-training framework, aimed at alleviating the performance constraints encountered in IA segmentation due to the limited availability of annotated DSA data. Our extensive experiments on the DIAS dataset demonstrate the effectiveness of these methods as potential baselines for future research and clinical applications. The dataset and code are publicly available at \url{https://doi.org/10.5281/zenodo.11396520} and \url{https://github.com/lseventeen/DIAS}.
\end{abstract}

\section{Introduction}
\label{sec:introduction}

\IEEEPARstart {C}{erebrovascular} diseases substantially contribute to global mortality and long-term disability~\cite{2022global}. Recent decades have witnessed a decline in mortality from Intracranial Artery (IA) diseases like Intracranial Arterial Stenosis (ICAS), Middle Cerebral Artery Occlusion (MCAO), and intracranial aneurysm, thanks to advancements in medical and imaging technology~\cite{imaging}. Among these, Digital Subtraction Angiography (DSA), an imaging modality, visualizes the flow of a contrast agent through vessels by capturing a sequence of images over a duration typically spanning 3 to 15 seconds, with a sampling rate between 3 to 5 frames per second~\cite{fully}. It effectively removes bony structures by subtracting pre-contrast images, thereby enhancing the visibility of vessels filled with the contrast agent. Owing to its inherently superior spatial and temporal resolution, DSA can accurately reveal lesion details in instances where Computed Tomography Angiography (CTA) and Magnetic Resonance Angiography (MRA) may not offer a definitive diagnosis~\cite{imaging}. Thus, DSA is universally acknowledged as the gold standard for investigating lesion angioarchitecture, deciphering the dynamics of arterial blood supply, and guiding endovascular intervention~\cite{dsa2022}. 

\begin{figure}[tbp]
\centering
\includegraphics[scale=0.55]{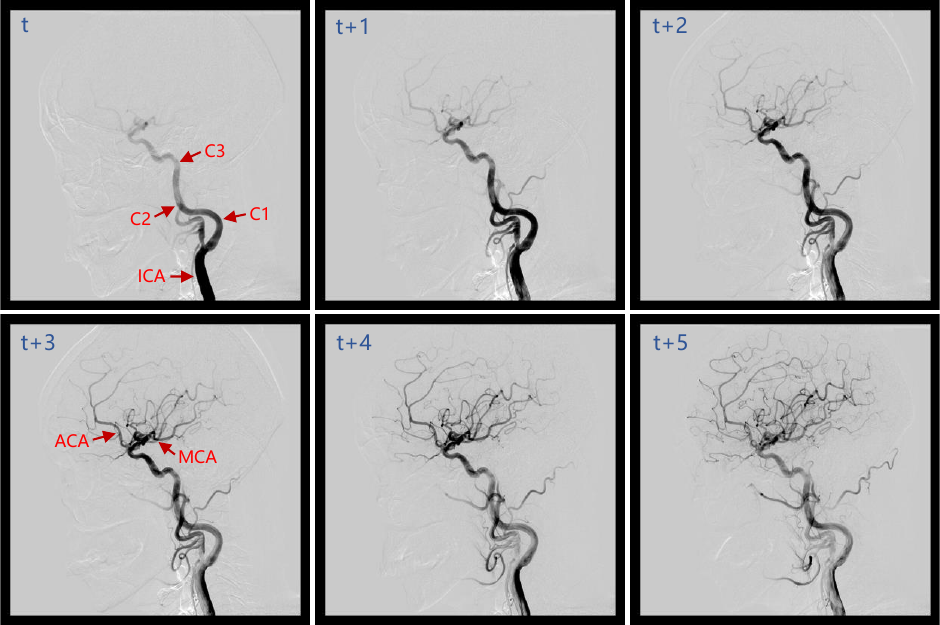}
\caption{A representative sample of intracranial vessels in lateral view from DSA, where t to t+5 are consecutive frames of the arterial phase. The red arrow in the upper left image indicates segments C1-C3 of the intracranial internal carotid artery (ICA), while the lower left image represents the anterior cerebral artery (ACA) and the middle cerebral artery (MCA).} \label{dsa_eg}
\end{figure}

Conventionally, neuroradiologists analyze DSA images visually for diagnosis, clinical decision-making, and postoperative evaluation, which is inherently time-consuming, labor-intensive, and subjective. Automated vessel segmentation in DSA promises to alleviate these challenges by accurately outlining vascular morphology. Currently, DSA-based cerebrovascular image segmentation finds utility in various downstream clinical applications, including automated thrombolysis in cerebral infarction scoring~\cite{tici}, 3D reconstruction of vessel~\cite{2d-3d-segmentation}, and image guidance of interventional surgery~\cite{spiegel20112d}. Despite its critical importance, cerebrovascular segmentation in DSA has not been extensively researched. A contributing factor to this issue is the lack of a dedicated dataset designed for the development and evaluation of vessel segmentation algorithms in DSA. However, the creation of such a dataset is particularly challenging due to the unique characteristics of DSA imaging. Primarily, in the DSA sequence, the intravascular contrast agent is projected from a fixed angle, revealing only a portion of the intracranial arteries within the contrast medium in each image, as illustrated in Fig.~\ref{dsa_eg}. Annotating vessels in DSA necessitates neuroradiologists to meticulously examine each image frame, a process that is both labor-intensive and costly~\cite{su2022spatio}. Moreover, the intricate nature of cerebral vessels, characterized by overlapping structures, reduced contrast in fine vessels, and complex background noise, significantly increases the difficulty of accurate annotation.

Currently, several studies~\cite{meng2020multiscale, li2020cau, su2022spatio,weakly} have emerged in the domain of vessel segmentation within DSA, demonstrating exceptional performance on in-house datasets. Nevertheless, the majority of these methods~\cite{meng2020multiscale, li2020cau,weakly} focus on segmenting single-frame images, thereby neglecting the inherent temporal attributes of DSA sequences. Constrained by the fact that single-frame DSA captures only a fraction of the vascular contrast, such methods may inadequately represent the complex vascular structure of IA. This shortfall could potentially undermine their utility in the accurate diagnosis and treatment of cerebrovascular diseases~\cite{su2022spatio}. Vessel segmentation in DSA should be perceived as a dimensionality reduction task, where the input is 2D+Time images, and the output is a 2D segmentation map. This approach deviates from the traditional 2D and 3D vessel segmentation methods~\cite{phtrans,fr-unet,maa-unet}. The key challenge lies in accurately capturing the spatiotemporal information present within the angiography for precise vessel segmentation.

\begin{table*}[tbp]
\centering
\caption{Summary of several public vessel segmentation datasets. $^\dagger$ represents only training set data is available.} \label{table:dataset}
\renewcommand\arraystretch{1.5}
\tabcolsep5pt
\scriptsize
\begin{tabular}{lcccccc}
\hline
Dataset & Year   & Target  & Modality     & Dimension & Numbers & Resolution  \\ \hline
DRIVE~\cite{DRIVE} & 2004 & retinal vessels              & color fundus & 2D             & 40     & 565×584     \\
CHASE\_DB1~\cite{CHASEDB}& 2012 & retinal vessels              & color fundus & 2D             & 28     & 999$\times$960     \\
STARE~\cite{STARE} & 2000     & retinal vessels              & color fundus & 2D             & 20     & 700$\times$605   \\
HRF~\cite{HRF} & 2013     & retinal vessels              & color fundus & 2D             & 45     & 3504$\times$2336 \\
ROSE1~\cite{ROSE} & 2020    & retinal vessels              & OCTA         & 2D             & 117    & 304$\times$304     \\
ROSE2~\cite{ROSE} & 2020     & retinal vessels              & OCTA         & 2D             & 112    & 512$\times$512     \\
CHUAC~\cite{CHUAC} & 2018     & coronary artery     & angiograpy   & 2D             & 30     & 189$\times$189     \\
DCA1~\cite{DCA1} & 2019      & coronary artery     & angiograpy   & 2D             & 134    & 300$\times$300     \\
ASOCA$^\dagger$~\cite{ASOCA} & 2022    & coronary artery     & CTA          & 3D             & 40     & 512$\times$512$\times$200 \\ 
VESSEL12$^\dagger$~\cite{VESSEL12}& 2014  & lung vessel         & CT           & 3D             & 20     & -           \\
\textbf{DIAS}& 2023 & intracranial artery & DSA          & 2D+Time        & 120 (753)    & 800$\times$800$\times$4-14  \\ \hline
\end{tabular}
\end{table*}

In this work, we propose the \textbf{D}SA-sequence \textbf{I}ntracranial \textbf{A}rtery \textbf{S}egmentation dataset (\textbf{DIAS}), meticulously annotated at the pixel level by neurosurgeons. Acknowledging the complexity and labor-intensiveness of vessel annotation in DSA sequences, we've  provided two types of weak annotations for training samples to support research in weakly-supervised IA segmentation. To comprehensively evaluate the efficacy of prevalent segmentation methods, including fully-supervised, weakly-supervised, and semi-supervised techniques, on the DIAS dataset, we have established a benchmark comprising three key components:

\begin{itemize}[]
\item We propose the vessel sequence segmentation network (VSS-Net), incorporating the sequence feature extraction module adept at efficiently capturing spatiotemporal representations of intravascular contrast. Notably, VSS-Net is designed to be compatible with any encoder-decoder network, enabling precise segmentation of intracranial arteries in 2D+Time DSA sequences.

\item We propose an innovative scribble learning-based image segmentation framework, specifically designed for weakly-supervised IA segmentation. This framework, distinguished by its dual-model architecture, employs cross pseudo-supervision and consistency regularization, guided by scribble labels. This integration significantly boosts the performance of the segmentation network, capitalizing on the synergistic advantages offered by the dual-model approach.
\item We introduce the random patch-based self-training algorithm, which alleviates the challenges arising from the limited availability of annotated DSA data in IA segmentation by enhancing the model's generalization through joint training on randomly selected image patches from both labeled and pseudo-labeled data.

\end{itemize}

Through extensive experimentation on the DIAS dataset, we have established robust baselines that hold the potential to advance future algorithmic developments in cerebral vessel segmentation.


\section{Related work}
\subsection{Vessel Segmentation Datasets}

Vessel segmentation, a prominent research field in computer-aided medicine, has seen significant advancements recently, largely due to technological progress and the availability of annotated datasets. We have summarized the several publicly available vessel segmentation datasets in Table~\ref{table:dataset}. A representative example is retinal vessel segmentation, which has many publicly available datasets, such as colour fundus datasets (DRIVE~\cite{DRIVE}, CHASE\_DB1~\cite{CHASEDB}, STARE~\cite{STARE}, and HRF~\cite{HRF}) and OCTA datasets (ROSE1 and ROSE2~\cite{ROSE}). These datasets consist of 20-117 2D images with spatial resolution ranging from 304$\times$304 to 3504$\times$2336. However, other vessel segmentation datasets are relatively scarce~\cite{su2022spatio}. CHUAC~\cite{CHUAC} and DCA1~\cite{DCA1} are coronary angiography datasets with resolutions of 189$\times$189 and 300$\times$300, respectively. ASOCA~\cite{ASOCA} and VESSEL12~\cite{VESSEL12} are the 3D vessel segmentation datasets of coronary artery CTA and lung CT from the challenges, in which only training set data is available. 

The proliferation of such datasets has catalyzed a wave of innovative research in the area of segmentation, with scholars endeavoring to devise algorithms that surpass current methodologies. Until now, the medical imaging community lacked publicly available DSA datasets with vessel annotations, with only limited studies using proprietary data. DCVessel~\cite{meng2020multiscale} comprises 30 DSA arterial-phase images at a resolution of 700$\times$700, while HEART~\cite{BEFD} comprises 1092 DSA coronary images at 256$\times$256 resolution. Su \emph{et al.}~\cite{su2022spatio} manually segmented arteries and veins on 97 DSA series, covering both anteroposterior and lateral views, each with a resolution of 1024$\times$1024 pixels. Our contribution, DIAS, offers a comprehensive IA segmentation dataset from DSA sequences, comprising 120 sequences (753 frames) at 800$\times$800 pixel resolution, spanning lengths of 4 to 14 frames. The DIAS can serve as a reliable benchmark for the development of algorithms on IA segmentation, hopefully inspiring further research in this arena.
\subsection{Vessel Segmentation Methods}
In the past few years, deep learning has revolutionized vessel segmentation, outperforming traditional approaches like hand-crafted features~\cite{zhang2018vesselness}, filtering-based models~\cite{sang2007knowledge}, and statistical models~\cite{liu2018vessel}. Deep learning-based methods, particularly UNet~\cite{unet} and its numerous variants, have become the primary methods for vessel segmentation. These networks typically employ an encoder-decoder architecture with popular components, such as multi-scale design~\cite{maa-unet,meng2020multiscale,scsnet}, deep supervision~\cite{wu2019vessel,meng2020multiscale,fr-unet}, attention mechanisms~\cite{cs2-net,maa-unet,agnet}, atrous convolution~\cite{rvsegnet,wu2019vessel,fr-unet,meng2020multiscale}, and improved skip connections~\cite{meng2020multiscale,fr-unet}. Moreover, several novel neural networks~\cite{rvgan,gnn} have also succeeded in vessel segmentation by leveraging generative adversarial networks (GAN) or graph neural networks (GNN). Beyond network architectures, some research has focused on optimizing training strategies~\cite{sgl,boosting} and modifying loss functions~\cite{loss}.

There have been few studies on DSA vessel segmentation. \cite{meng2020multiscale} proposed a CNN-based multiscale dense CNN segmentation network for IA segmentation in single-frame DSA, utilizing multiscale atrous convolution, enhanced dense blocks, and redesigned skip connections. \cite{BEFD} proposed a boundary enhancement and feature denoising module to improve the extraction of boundary information in DSA coronary segmentation. These two approaches are tailored for vessel segmentation in individual DSA frames. For vessel sequence, \cite{svsnet} proposed the sequential vessel segmentation network (SVS-Net) for vessel segmentation in X-ray coronary angiography, which incorporates 3D convolutional layers to extract rich temporal-spatial features and employs channel attention blocks to learn discriminative features. \cite{su2022spatio} proposed a spatio-temporal U-Net (ST U-Net) architecture which integrates convolutional gated recurrent units (ConvGRU) in the contracting branch of U-Net for cerebral artery and vein segmentation in DSA sequences. While both methods achieve vessel segmentation within sequences, the former is not adaptable to variable sequences, and the latter is challenging to extend to other models.


Fully annotating vessel datasets is both time-consuming and costly. Consequently, annotation-efficient segmentation methods based on weakly-supervised and semi-supervised have attracted significant attention from researchers. Weakly-Supervised Segmentation (WSS) trains models with sparse or noisy annotations, instead of exact and complete annotations. The common methods involves online labeling~\cite{chen2021seminar} via consistency checks with the mean teacher or using pseudo-labels supervision~\cite{luo2022scribble} derived from model predictions. Furthermore, investigating the method for weakly-supervised annotation of vessels in DSA is a research direction deserving of exploration. \cite{weakly} propose a novel learning framework, which utilizes an active contour model for weak supervision and low-cost human-in-the-loop strategies to improve weak label quality. Semi-Supervised Segmentation (SSS) aims to explore tremendous unlabeled data with supervision from limited labeled data. Recently, self-training methods~\cite{st++,flare22} have dominated this field. Furthermore, methods employing consistency regularization strategies~\cite{st++,cps,ouali2020semi} improve the generalization ability by encouraging high similarity in predictions from two perturbed networks for the same input image.

Various domain adaptation techniques have been proposed to enhance the generalizability of vessel segmentation, leveraging vessel data from other modalities with similar anatomical structures to aid in training, thereby improving the performance of models in SSS. \cite{gu2022contrastive} propose contrastive SSS for cross anatomy domain adaptation that adapts a model to segment coronary arteries, which requires only limited annotations in the coronary arteries by leveraging existing annotated retinal vessels of similar structure. \cite{a2v} propose a semi-supervised domain adaptation framework for accurately segmenting 3D brain vessels by employing the StyleGAN2~\cite{stylegan2} to facilitate the representation of heterogeneous volumetric data and bridge the substantial domain gap between angiography and venography brain images. To our knowledge, there is no existing literature on weakly-supervised and semi-supervised vessel segmentation in DSA sequences. Building upon our created dataset, we have conducted in-depth research and proposed innovative methods in three distinct areas: network models, WSS, and SSS. These approaches are specifically tailored to leverage the unique features of the dataset, addressing each segmentation paradigm comprehensively.

Furthermore, large-scale pretrained models, i.e., foundation models, exhibit considerable generalizability and adaptability, positioning them as a focal point of recent medical image analysis research~\cite{DBLP:conf/iccv/ButoiOMSGD23}. However, generally, a unified foundation model approach cannot achieve state-of-the-art performance in many medical image analysis tasks due to large variations present in organs and important structures, texture, shape, size and topology (e.g., blood vessels), and imaging modalities~\cite{zhang2023challenges}. We are convinced that, through continuous research, foundation models possess the potential to achieve vascular segmentation in DSA that can compete with fully supervised methods, leveraging few-shot learning, zero-shot learning, or prompt engineering.


\section{DIAS Dataset}

\subsection{Dataset Description}
The DIAS dataset is derived from DSA images obtained during ischemic stroke treatments at Beijing Tiantan Hospital. This dataset encompasses cases of ICAS and MCAO. The ethical committee at Beijing Tiantan Hospital authorized the use of all the data presented in this study. All images were retrospectively collected between January 2019 and December 2021 and then anonymized to protect the patients' privacy by removing all identifiable clinical treatment details. The original DSA sequences were captured by the Siemens angiography machine during the neurointerventional procedures and stored as DICOM files. It contains 1027 2D-DSA sequences from 416 patients in anteroposterior view or lateral view. In this study, we retained 120 sequences (762 images) after excluding those with significant motion artifacts or incomplete arterial phase, or any duplicates. A total of 120 sequences were derived from 60 patients, among whom 24 were female and 36 were male. The average age of these patients was 62.5 years, with an age range from 40 to 88 years old. The number of frames varies between sequences (6 - 46 frames), but the frame rate is fixed (4 FPS). The sequences were then extracted into individual images with a resolution of 800$\times$800 pixels using RadiAnt. We retained all the arterial-phase frames and removed the pre-contrast, capillary-phase, and venous-phase frames under the guidance of a neurosurgeon. As a result, the length of DSA sequences in the DIAS dataset ranges from 4 to 14 frames.


\begin{figure}[tbp]
\centering
\includegraphics[scale=0.32]{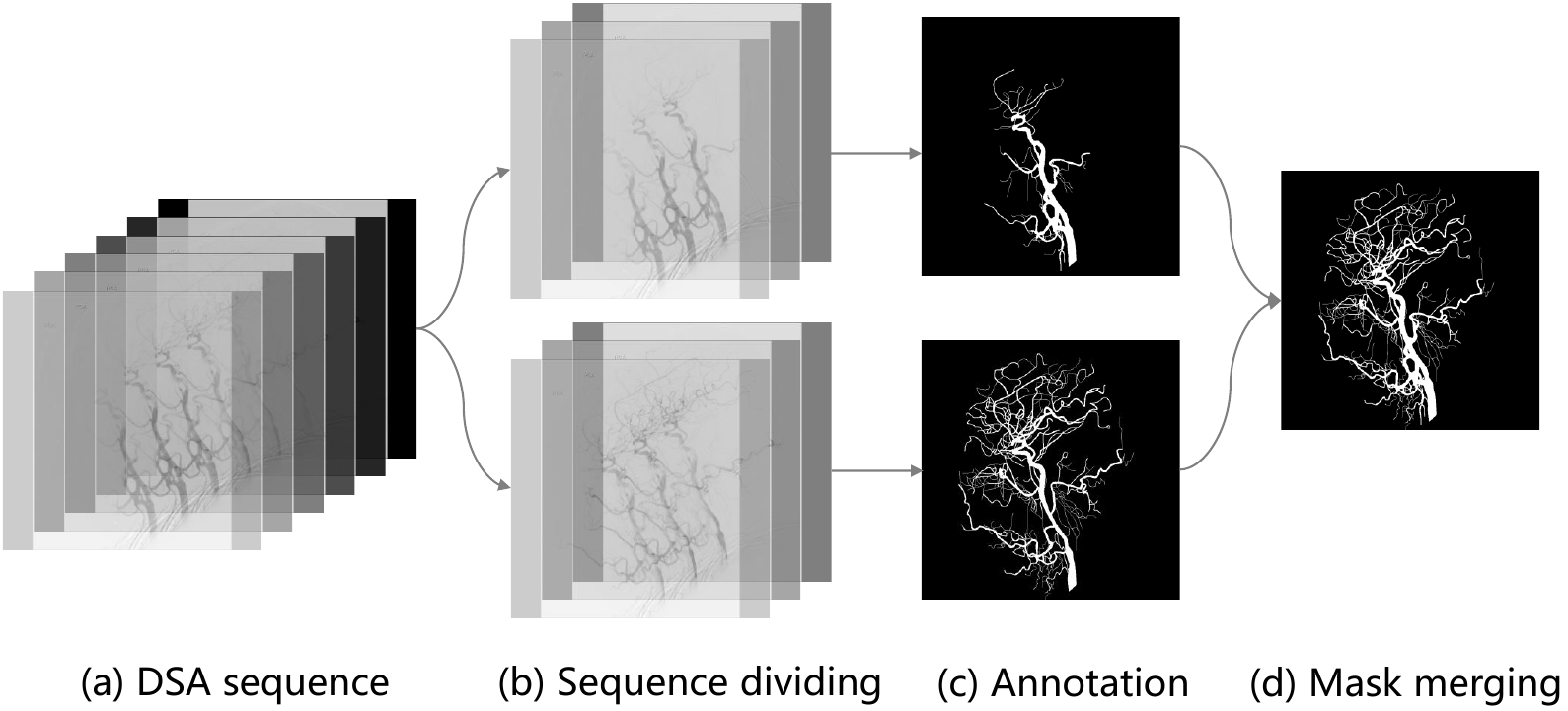}
\caption{Full annotated process of DIAS.} \label{annotated_process}
\end{figure}

\subsection{Reference Annotation Protocol}
\subsubsection{Full Annotation}

The manual annotation of IA in DSA sequences necessitates a thorough consideration of all image data within the sequence. Fig.~\ref{dsa_eg} illustrates a typical representation of intracranial vessels across five successive frames during the arterial phase in a lateral view. During this phase, the contrast agent progressively spreads from the Internal Carotid Artery (ICA) into cerebral vessels like the Anterior Cerebral Artery (ACA) and Middle Cerebral Artery (MCA), carried by the blood flow. For a single image, the contrast agent is discernible in only a portion of the IA. For instance, the contrast agent at t filled the C1-C3 segment of the ICA, while at t+5, it filled the ACA and MCA. To annotate the IA effectively, each image in the sequence must be individually annotated, and then these annotations should be integrated to leverage the full vessel information available in the sequence, thereby ensuring precision in annotation. However, the challenge arises with the fine nature of vessels, especially in images from later arterial stages, which makes the standard approach of demarcating their borders impractical for medical image annotation. Thus, we have to annotate each vessel pixel by pixel, which is a labor-intensive and time-consuming procedure. In reality, the angiographic representations of vessels in adjacent frame images exhibit a high degree of similarity, indicating that annotating every pixel in each frame image involves a substantial amount of repetitive work. Fig.~\ref{annotated_process} presents our proposed solution, a pragmatic compromise that aims to reduce the workload without compromising the accuracy of annotations:

\begin{figure}[tbp]
\centering
\includegraphics[scale=0.34]{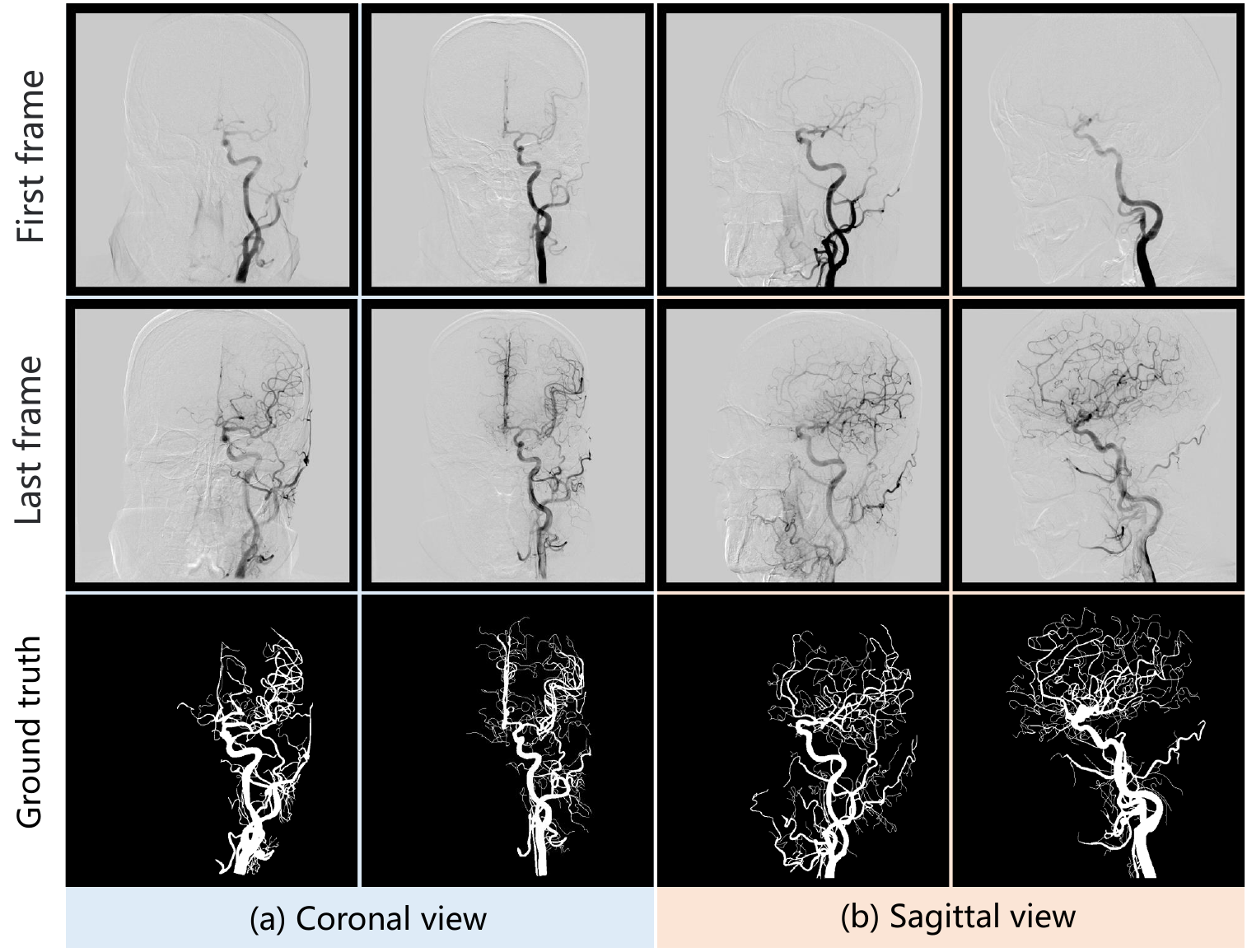}
\caption{Full annotated representative samples of DIAS at anteroposterior view and lateral view.} \label{label_show}
\end{figure}

\begin{figure}[tbp]
\centering
\includegraphics[scale=0.47]{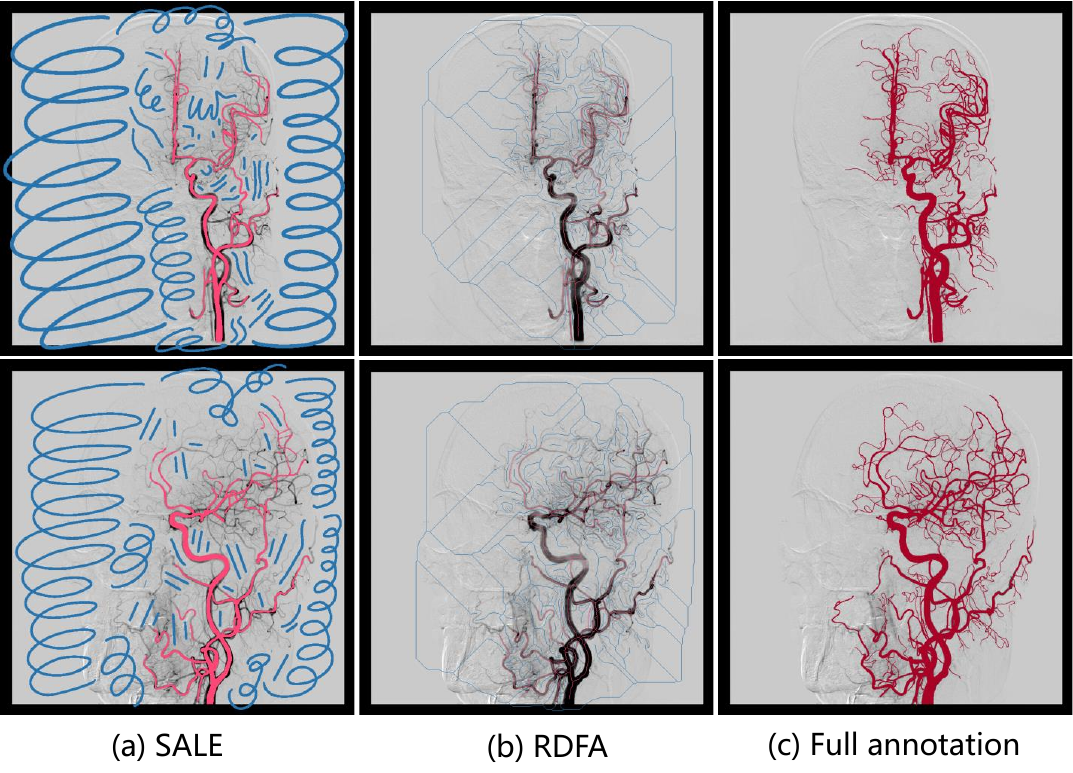}
\caption{Examples of two forms scribble annotations: the Scribbles Annotation with Little clinical Experience (SALE) and Random Drawing based the Full Annotation (RDFA). Red represents annotated vessel pixels, blue represents annotated background pixels.} \label{scribble_show}
\end{figure}

\begin{itemize}[]
\item
The arterial-phase DSA sequences were divided into two phases from the middle: the anterior phase and the posterior phase. The contrast agent during the anterior phase predominantly fills the ICA as opposed to the smaller vessels such as the ACA and MCA arteries, which are filled by the contrast agent in the posterior phase.
\item
The last image of each phase is annotated pixel-wise, followed by a meticulous review of other same-phase images to finalize the annotation. Thick vessels are the focus of annotation in the anterior phase, while thin vessels are prioritized in the later phase.
\item
The annotations of thick and thin vessels are merged using a pixel-level OR operation (1 for vessel pixel and 0 for background) to create the ultimate groudtruth.
\end{itemize}

We randomly selected 60 sequences from the DIAS dataset for the purpose of annotating vessels, using patient IDs as the criterion for selection. The remaining 60 sequences were used for research on semi-supervised segmentation of IA. The DSA sequence was initially annotated by two MD students specializing in neurosurgery, followed by corrections from an associate physician with 10 years of experience, and finally confirmed and double-checked by a chief physician with 15 years of experience. Fig.~\ref{label_show} presents fully annotated examples of intracranial arteries in anteroposterior and lateral views.

\begin{figure*}[h]
\centering
\includegraphics[scale=0.56]{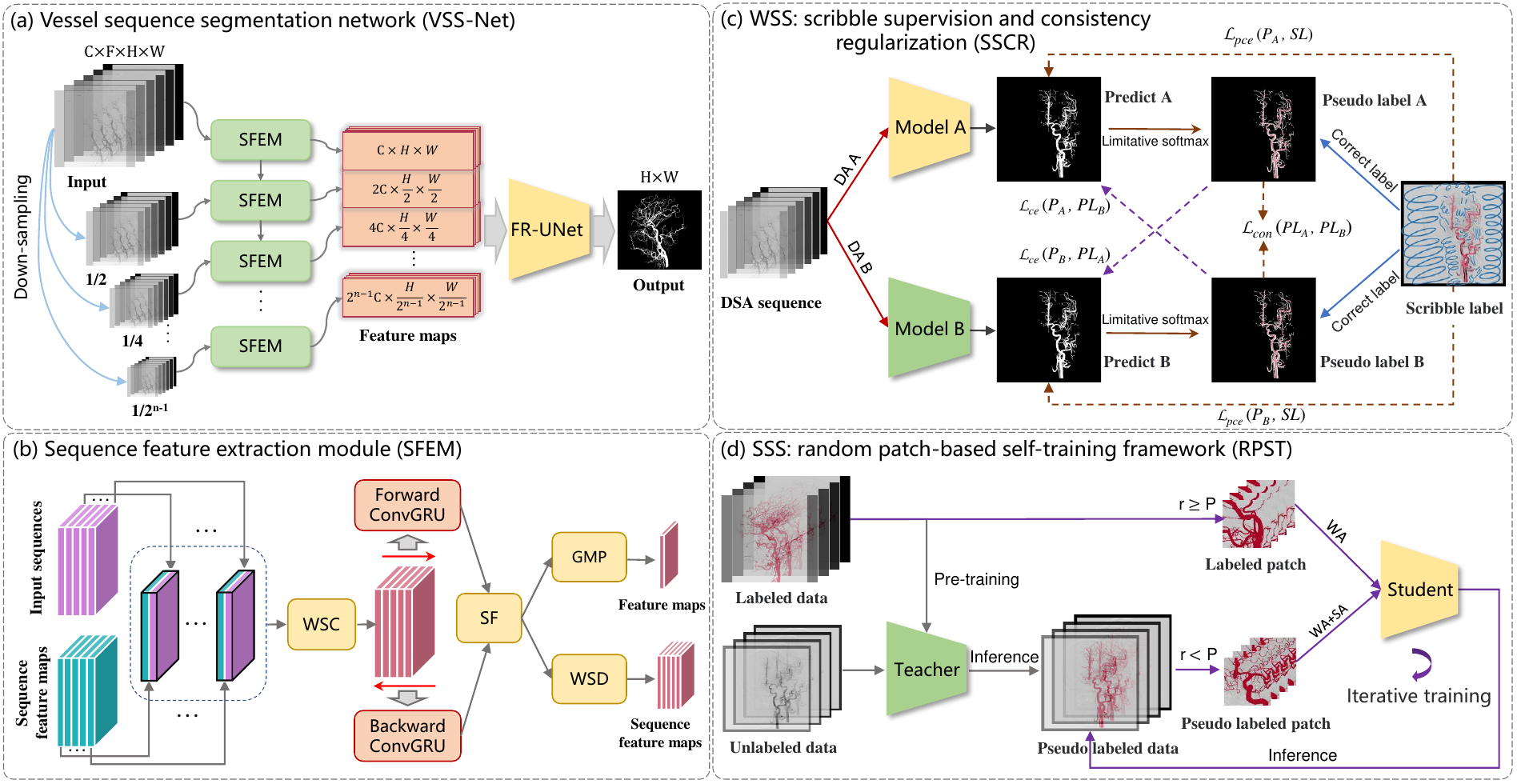}
\caption{Illustration of the proposed benchmark for DSA-sequence intracranial artery segmentation. (a) Vessel sequence segmentation network with (b) sequence feature extraction module; (c) WSS: scribble learning-based segmentation composed of scribble supervision and consistency regularization; (d) SSS: random patch-based self-training framework. (WSC: Weight Sharing Convolution, SF: Sequence Fusion, GMP: Global Max Pooling, WSD: Weight Sharing Downsampling, DA: Data augmentation, WA: Weak Augmentation, SA: Strong Augmentation) } \label{method}
\end{figure*}

\subsubsection{Scribble Annotation}

Scribble-based image segmentation has proven to be an efficacious and sought-after method for WSS~\cite{rloss}. This approach to weak annotation allows annotators to easily mark object centers without the challenging task of delineating class boundaries. To explore the potential of scribble supervision for IA segmentation in DSA sequences, we generated two forms of scribble annotations for the training set of DIAS: the Scribbles Annotation with Little Clinical Experience (SALE) and Random Drawing based on the Full Annotation (RDFA). SALE simulates vessel annotation with little expert experience, wherein the thick vessels highlighted by the contrast agent are annotated, as illustrated by the red marker in Fig.~\ref{scribble_show} (a). Despite receiving clinicians’ basic training, accurately distinguishing the thin vessels and edges of thick blood vessels with low contrast remains challenging. Consequently, pixels in these regions were not annotated during annotation. Moreover, scribble segmentation relies on supervised information for every category. Apart from annotating the vessels, we conducted extensive scribble annotations for the background wherever possible, as illustrated by the blue marker in Fig.~\ref{scribble_show} (a). Follow~\cite{ivm}, we generate RDFA by skeletonizing and randomly pruning each category of the full annotation, as shown in Fig.~\ref{scribble_show} (b). RDFA simulates random scribble annotations, diverging from SALE, which focuses on easier annotations. In SALE and RDFA, we set the pixel values of the background annotation and vessel annotation to 0 and 1, respectively, while pixels without annotations are assigned a value of 255.


\section{Methods}

Automatic IA segmentation in DSA sequences is an under-explored and challenging task. To fully explore potential research topics, an elaborately designed benchmark for IA segmentation was proposed, aiming to provide comprehensive evaluation of the DIAS dataset using Fully-Supervised Segmentation (FSS), WSS, and SSS, which are increasingly drawing attention in the medical image analysis community.

\subsection{Fully-supervised Segmentation}
Segmenting IA in DSA sequence can be considered as a Dimensionality Reduction (DR) task. Given an image $I \in \mathbb{R}^{\prod_{d=1}^{M}n_d}$, where $M$ represents the number of dimensions and $n_d$ is the size of the image in the corresponding dimension $d$, DR segmentation can be defined as finding a function $f$ such that $f: I \rightarrow O$, where $O \in \mathbb{R}^{\prod_{d=1}^{N}n_d}$ represents segmentation output and $N $ ($<M$) is its dimensionality~\cite{PSC}. For IA segmentation in DSA sequence, $M=3$, $N=2$. 

Contrary to traditional medical image segmentation models that solely focus on spatial features, our proposed vessel sequence segmentation network efficiently captures spatio-temporal representations of vessels across various scales to achieve precise IA segmentation in DSA. Figure~\ref{method} (a,b) illustrates an overview of the proposed network architecture. In VSS-Net, we employ Sequence Feature Extraction Modules (SFEM) at each stage in the encoder to extract feature maps across multiple scales. These modules handle the 2D+Time input sequences and incorporate feature maps from the preceding stage, resulting in 2D feature maps that capture both spatial and temporal representations. With the deepening of stages, the input resolution is reduced, consequently expanding the receptive field that facilitates the generation of high-dimensional spatiotemporal representations. Subsequently, these feature maps can be fed into any 2D model following the encoder-decoder framework, enabling vessel segmentation in the sequence.

In the encoder, the input feature map from each stage is integrated with the raw sequence, serving to supplement fine-grained semantic information, except for the initial stage. Specifically, the raw sequence has been resized to be consistent with the corresponding feature maps by bilinear interpolation. Each image in the sequence is concatenated with feature maps from the same temporal sequence, and then they are fused using a 1$\times$1 convolution with shared weights (WSC), leading to a reduction by half in the number of channels. 
The calculation process for the fusion output $Q^{j}$ at scale 
$j$ is as follows:
\begin{equation}
Q^{j} = \sum_{i=1}^{N} W \ast(x^j_i \oplus f^j_i), \quad \text{for } j \in \left\{ \frac{1}{2}, \frac{1}{4},  \ldots, \frac{1}{2^{n-1}} \right\},
\end{equation}
where $x^j_i$ represents the $i^{th}$ frame of the input sequence, scaled to $j$ times the original resolution, $f$ is the feature map, $N$ denotes the length of the sequence, $W$ are $1\times1$ convolutional kernels, $\ast$ denotes the convolution operation, $\oplus$ denotes the concatenation operation, $n$ denotes the number of stages in the encoder.

Subsequently, we employ a bidirectional ConvGRU to construct the spatiotemporal connection between the feature maps of the sequence. ConvGRU~\cite{convgru} is an extension of the traditional GRU that integrates convolutional operations, making it more suitable for spatial data like images. Consider $x_t$ as the input to the ConvGRU, with corresponding hidden states denoted by $h_t$, we can delineate the updating process of ConvGRU unrolled by time as follows:
\begin{align}
z_t &= \sigma(W_z \ast x_z + U_z \ast h_{t-1}), \\
r_t &= \sigma(W_r \ast x_r + U_r \ast h_{t-1}), \\
\tilde{h}_t &= \tanh(W_h \ast x_t + U_h \ast (r_t \odot h_{t-1})), \\
h_t &= (1 - z_t) \odot h_{t-1} + z_t \odot \tilde{h}_t,
\end{align}
where $\sigma$ represents the sigmoid function, $\tanh$ represents the hyperbolic tangent function, $\odot$ denotes the Hadamard product, i.e., element-wise multiplication, $W_r$, $W_z$, $W_h$ and $U_r$, $U_z$, $U_h$ are 2D-convolutional kernels.

We simultaneously capture the spatiotemporal dynamics of vascular flow in both forward and backward directions. Subsequently, the outputs of the bidirectional ConvGRU are fused into a single sequence~\cite{DB-convgru}. The resulting integrated features encompass angiographic representations of both macrovascular and microvascular structures, spanning the early to late arterial stages. The process can be formulated as follows:
\begin{align}
h_t^f &= ConvGRU(h_{t-1},x_t),\\
h_t^b &= ConvGRU(h_{t+1},x_t),\\
h_t &= \tanh(W_f \ast h_t^f + W_b \ast h_t^b),
\end{align}
where $h_t^f$ and $h_t^b$ denote the hidden states from the forward and backward ConvGRU units, respectively. $x_t$ is the $t^{th}$ output feature from WSC.

The output feature maps are spatially downsampled using stride convolution with weight sharing to serve as inputs for the next stage encoder. Simultaneously, we employ global max pooling along the sequence dimension of the feature maps for dimensionality reduction. The generated multi-scale feature maps are compatible with any U-shaped segmentation architecture. In particular, we integrate it with FR-UNet, where the outputs of the multi-scale SFEM are fed into each stage of FR-UNet. The resultant VSS-Net effectively harnesses the strengths of both components, extracting spatiotemporal representations to realize precise vessel segmentation within sequences.

\subsection{Weakly-supervised Segmentation}
\label{Weak-supervised}

A prevalent strategy for scribble-supervised semantic segmentation is to handle the problem as a fully supervised task, using a loss function $\mathcal{L}$ on the available labeled pixels. However, this baseline approach fails to harness the unlabeled points that contain critical semantic information. Consequently, the absence of sufficient labeled data invariably results in performance degradation. To enhance the segmentation performance of scribble, the key lies in extending the supervision of labeled pixels to unlabeled pixels. To overcome these issues, we propose SSCR, a novel framework for learning-based scribble image segmentation composed of scribble supervision and consistency regularization. Fig.~\ref{method} (c) presents the implementation framework for SSCR. SSCR generates pseudo-annotations for the unlabeled pixels with guidance provided by scribble supervision, utilizing partial cross-entropy to extend supervised information, and maximizes the performance of scribble segmentation by regularizing the segmentation network using mutual consistency and cross pseudo supervision.

We define $P$ as the set of image pixels and $S \subseteq P $ as the set of annotated pixels. The proposed approach feeds $P$ into two segmentation networks with different augmentation strategies. The outputs of each network are supervised by applying the cross-entropy function solely to the annotated pixels on the scribble labels. The function, namely partial cross entropy (pCE), can be formulated as:
\begin{equation}
\mathcal{L}_{pce}(y,\hat{y})=-\sum_{i=1}^M\sum_{j=1}^N{ \mathbbm{1}(p_{i,j}\in S)(y_{i,j}\log(\hat{y}_{i,j|\theta})}),
\end{equation}
where $M$ and $N$ represent the width and height of the image, respectively. $\hat{y}_{i,j|\theta}$ is the predict for the pixel $p_{i,j}\in S$ given the network parameters $\theta$, and $y_{i,j}$ denotes the ground truth label. 

To generate pseudo labels, we first calculate the class distribution of the one-hot prediction map using limited softmax, which exclusively calculates the channel corresponding to the annotated class. Then, we apply scribble annotations to correct the pseudo labels in order to improve their confidence. The formula for creating the pseudo label is as follows:

\begin{equation}
PL_i=
\begin{cases}

s_{i,j},&\text{if}\ p_{i,j} \in S  \\
\text{argmax}(\text{softmax}(\mathcal{H}(p_{i,j};\theta)[C])), &\text{otherwise} 

\end{cases}
\end{equation}
where $s_{i,j} \in S$ defines annotated pixels, and $\mathcal{H}(p_{i,j};\theta)$ defines the predicted output of $p_{i,j}$. $C$ is the channel of annotated class, i.e., 0 and 1.

We enforce network consistency by utilizing cross pseudo supervision, which employs pseudo labels from one network to supervise the other network~\cite{cps}. Additionally, these pseudo labels are imposed a consistency loss to penalize inconsistent segmentation. The training objective contains three losses: scribble supervision loss $\mathcal{L}_{scr}$, cross pseudo supervision loss $\mathcal{L}_{cps}$, and consistency loss $\mathcal{L}_{con}$:
\begin{align}
\mathcal{L}_{scr}=&\mathcal{L}_{pce}(P_A,SL)+\mathcal{L}_{pce}(P_B,SL) ,\\
\mathcal{L}_{cps}=&\mathcal{L}_{ce}(P_A,PL_B)+\mathcal{L}_{ce}(P_B,PL_A),\\
\mathcal{L}_{con}=&\mathcal{L}_{mse}(PL_A,PL_B),
\end{align}
where $\mathcal{L}_{ce}$ denotes the cross-entropy loss function and $\mathcal{L}_{mse}$ denotes the mean square error loss function, $P_A$ and $P_B$ are the outputs of two models, $PL_A$ and $PL_B$ are the pseudo labels, $SL$ is the scribble label. 

Finally, the training objective $\mathcal{L}$ is formulated as:
\begin{equation}
\mathcal{L}=\mathcal{L}_{scr} + \lambda_1\mathcal{L}_{cps} + \lambda_2\mathcal{L}_{con},
\end{equation}
where $\lambda_1$ and $\lambda_2$ are the weight of loss components $\mathcal{L}_{cps}$ and $\mathcal{L}_{con}$.

\subsection{Semi-supervised Segmentation}

Annotation of cerebral vessels manually from DSA image sequences is a laborious and time-consuming task. Semi-supervised learning has shown the potential to alleviate this problem by learning from a large set of unlabeled images and limited labeled samples~\cite{flare22}. The primary challenge in semi-supervised setting lies in how to effectively utilize the unlabeled images. Self-training~\cite{pseudo} via pseudo labeling is a conventional, simple, and popular pipeline to leverage unlabeled data, where the retrained student is supervised with hard labels produced by the teacher trained on labeled data. For semi-supervised segmentation problem, let $\mathcal{D}^{l}=\{(x_{i},y_{i})\}_{i=1}^{M}$ be pixel-wise labeled images and $\mathcal{D}^{u}=\{u_{i}\}_{i=1}^{N}$ be unlabeled images, where in most cases $ N \gg M$. Self-training can be simplified into the following steps: 
\begin{enumerate}[leftmargin=*,itemsep=-4pt]
\item Train a teacher model $T$ on labeled data $\mathcal{D}^{l}$; 
\item Predict hard pseudo labels on unlabeled data $\mathcal{D}^{u}$ with $T$ to obtain $\mathcal{\hat{D} }^{u}=\{(u_{i},T(u_{i})\}_{i=1}^{N}$;
\item Re-train a student model $S$ on the label data and pseudo-label data $\mathcal{D}^{l} \cup \mathcal{\hat{D} }^{u}$ for final evaluation.
\end{enumerate}

As shown in Fig.~\ref{method} (d), we follow the plainest self-training strategy to provide a SSS benchmark for DIAS. The injection of strong data augmentations on unlabeled images is highly advantageous as it decouples the predictions as well as alleviates overfitting on noisy pseudo-labels. Despite its simplicity, self-training with strong data augmentations surpasses existing methods without any bells and whistles~\cite{st++}. We performed two types of data augmentations in our implementation: weak and strong, denoted by $A^{w}$ and $A^{s}$ respectively. Weak Data Augmentations (WDA) consist of horizontal flipping, vertical flipping, and $[90, 180, 270]$ degree rotation. Strong Data Augmentations (SDA) include Cutout, adding white Gaussian noise, Gaussian blurring, and elastic deformation. Furthermore, we propose the application of random contrast adjustments to each image within the sequence to simulate angiography with intensity differences. We randomly extract patches from the DSA sequence to train the teacher and student models. Notably, the patches in each mini-batch were randomly extracted from $\mathcal{D}^{l}$ and $\mathcal{\hat{D}}^u$ with a probability $P$ when retraining the student model. Additionally, the final performance can be further improved by switching the teacher-student role and relabeling unlabeled images for iterative training. The pseudocode of the Random Patch-based self-Training framework, referred to as RPST, is provided in Algorithm~\ref{alg1}.

\renewcommand{\algorithmicrequire}{\textbf{Input:}}
\renewcommand{\algorithmicensure}{\textbf{Output:}}

\begin{algorithm}[h]
\caption{RPST}\label{alg1}
\KwIn{Labeled training set $\mathcal{D}^l = \{(x_i,y_i)\}_{i=1}^{M}$, \\    
\hspace{2.7em} Unlabeled training set $\mathcal{D}^u = \{u_i\}_{i=1}^{N}$, \\ 
\hspace{2.7em} Weak/strong augmentation $\mathcal{A}^w/\mathcal{A}^s$, \\ 
\hspace{2.7em} Teacher/student model $T/S$, \\
\hspace{2.7em} Execution probability $ P \in (0,1)$}  

\KwOut{Trained student model $S$}
Train $T$ on $\mathcal{D}^l$ with loss $\mathcal{L}$ \\
Obtain pseudo labeled $\mathcal{\hat{D}}^u = \{(u_i,T(u_i))\}_{i=1}^{N}$ \\
\For{Iterative training $N$ }{
    \For{Minibatch with the size of $B$ } {
        \For{$k \in \{1,...,B\}$}{
            \uIf{ Random number $ r > P$}{
            Randomly extract patch $(x^p_k,y^p_k)$ from $\mathcal{D}^l$ \\
                $x^p_k,y^p_k \leftarrow(\mathcal{A}^s\&\mathcal{A}^w)(x^p_k,y^p_k)$}
            \uElse {
            Randomly extract patch $(x^p_k,y^p_k)$ from $\mathcal{\hat{D}}^u$ \\ 
                $x^p_k,y^p_k \leftarrow\mathcal{A}^w(x^p_k,y^p_k)$}
        }
        $\hat{y}^p_k=S(x^p_k)$ \\
        Update $S$ to minimize $\mathcal{L}$ of $\{(\hat{y}^p_i,y^p_k)\}_{i=1}^{B}$
    }
    \uIf{Not the last iterative}{
    Obtain new pseudo labeled $\mathcal{\hat{D}}^u = \{(u_i,S(u_i))\}_{i=1}^{N}$ 
    }
    }
\end{algorithm}

\section{Experiments}
\subsection{Implementation Details and Evaluation Metrics}

The DIAS dataset was divided into 30 training samples, 10 validation samples, and 20 testing samples on the patient level. To facilitate input into the model, all DSA sequences were resampled to a length of 8. Furthermore, z-score normalization was applied across all DSA sequences for standardization. Patches of size $64\times64$ were randomly cropped for training, with a batch size of 64. The model parameters were optimized using AdamW \cite{adamw} with an initial learning rate of 5e-4. The learning rate is gradually reduced by the cosine annealing algorithm over 100 epochs. The parameter $P$ was set to 0.5 in the RPST, i.e., half of the patches in each batch were extracted from labeled data, while the rest came from unlabeled data in SSS experiment. For the SSCR, the weights of loss components $\lambda_1$ and $\lambda_2$ were empirically set to 1 and 0.5. We only use one network to generate the results for evaluation. All experiments used PyTorch with constant hyperparameters and were conducted on a single GeForce RTX 3090 GPU. We calculated the Dice Similarity Coefficient (DSC), the Area Under the receiver operating Characteristic curve (AUC), accuracy (Acc), sensitivity (Sen), specificity (Spe), and Intersection Over Union (IOU) to evaluate the performance of DSA sequence segmentation. Furthermore, we conducted evaluations of the Vascular Connectivity (VC) metric~\cite{fr-unet} and the connectivity-aware similarity measure (clDice)~\cite{cldice}.


\subsection{Fully-supervised Segmentation}
\label{sec:Fully-supervised Segmentation}

\begin{figure}[tbp]
\centering
\includegraphics[scale=0.54]{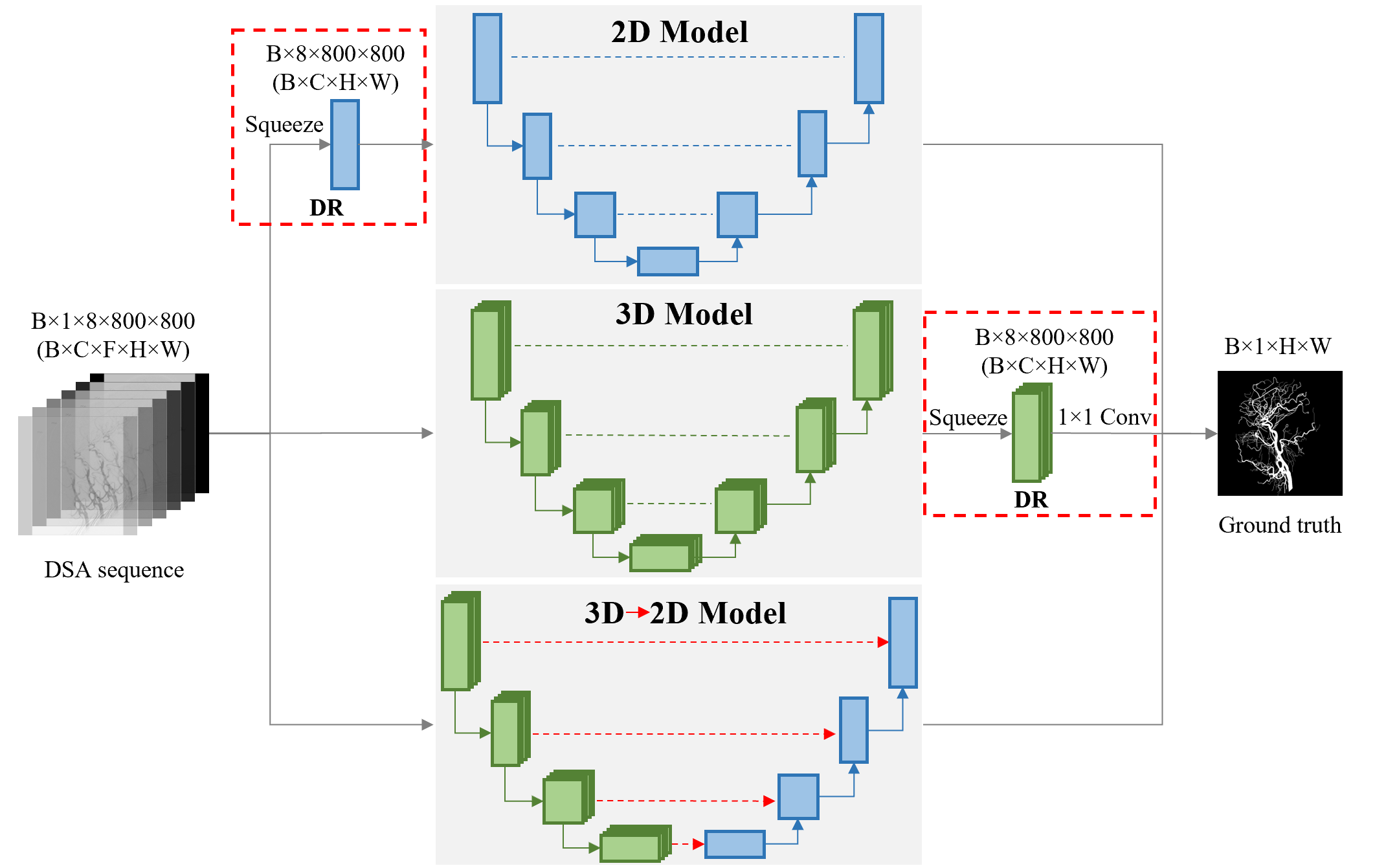}
\caption{Dimensionality reduction operation applied to 2D/3D models} \label{DR}
\end{figure}

We conducted comparative experiments on the DIAS using several state-of-the-art segmentation networks, which included traditional 2D/3D networks such as UNet~\cite{unet}, Attention-UNet~\cite{att-unet}, UNet++~\cite{unet++}, Res-UNet~\cite{res-unet}, MAA-Net~\cite{maa-unet}, CS$^{2}$-Net~\cite{cs2-net}, FR-UNet~\cite{fr-unet} and 3D$\rightarrow$2D networks like IPN~\cite{IPN}, PSC~\cite{PSC}, SVS-Net~\cite{svsnet}, ST U-Net~\cite{su2022spatio}. To order to achieve DR segmentation for these 2D/3D models, we propose a straightforward yet efficient DR operation that can convert them into DR models without altering their design. The DR operation, highlighted by the red box in Fig.~\ref{DR}, can be seamlessly incorporated as a plug-and-play module into either the input of a 2D model or the output of a 3D model. Specifically, given a DSA sequence $x \in \mathbb{R}^{C \times F \times H \times W}$, where $C$, $F$, $H$ and $W$ denote the number of channels, frame, height, and width, respectively. The initial number of channels is 1, which is squeezed by DR operation. The compressed input is then fed to the 2D model, where $F$ replaces $C$ as the input channel:
\begin{equation}
y = \mathcal{H}_{2d}(Squ(x)),
\end{equation}
where $Squ$ is dimensionality squeeze operation and $\mathcal{H}_{2d}$ is a 2D model. 
For a 3D model $\mathcal{H}_{3d}$, the DSA sequence is fed directly into the model, where it is treated as volume data. The dimension, $C$, of the model output is squeezed by DR operation. Finally, $1\times 1$ convolution reduces $F$ to the number of classes:
\begin{equation}
y = Conv_{1 \times 1}(Squ(\mathcal{H}_{3d}(x))).
\end{equation}

\begin{table*}[th]
\centering
\caption{Results of fully-supervised segmentation with state-of-the-art models. \textcolor{green!70!black}{Green}, \textcolor{blue}{blue}, and \textcolor{red}{red} respectively represent the highest scores for each metric in 2D, 3D, and 3D$\rightarrow$2D methods.} \label{table:model}
\renewcommand\arraystretch{1.5}
\tabcolsep3.2pt
\scriptsize

\begin{tabular}{clccccccccc}
\hline
\multicolumn{2}{c}{Methods}                           & \multirow{2}{*}{Params / M}    & \multicolumn{8}{c}{Metrics}                                                                                               \\ \cline{1-2} \cline{4-11} 
Type                   & \multicolumn{1}{c}{Networks} &         & DSC$\uparrow$      & Acc$\uparrow$       & Sen$\uparrow$     & Spe$\uparrow$        & IOU$\uparrow$     & AUC$\uparrow$        & clDice$\uparrow$         & VC$\downarrow$            \\ \hline
\multirow{7}{*}{2D}    & UNet                       & 31.04                      & 0.7700$\pm$0.0525 & 0.9654$\pm$0.0125 & 0.7570$\pm$0.0879 & 0.9839$\pm$0.0098 & 0.6290$\pm$0.0695 & 0.9829$\pm$0.0080 & 0.6939$\pm$0.0628 & 63.60$\pm$39.01  \\
                       & Res-UNet                  & 9.37 
                      & 0.7656$\pm$0.0548& 0.9655$\pm$0.0120& 0.7425$\pm$0.0950& 0.9850$\pm$0.0083& 0.6235$\pm$0.0719& 0.9816$\pm$0.0093& 0.6870$\pm$0.0745& 60.51$\pm$40.00  \\
                       & UNet++                      & 9.05                       & 0.7767$\pm$0.0468&0.9659$\pm$0.0122&\textcolor{green!70!black}{\textbf{0.7726$\pm$0.0752}} &0.9830$\pm$0.0091&0.6374$\pm$0.0630&0.9830$\pm$0.0084&0.7042$\pm$0.0589&54.79$\pm$32.24
   \\
                       & Att-UNet                 & 8.73       &                0.7760$\pm$0.0435&0.9660$\pm$0.0118&0.7686$\pm$0.0800&0.9835$\pm$0.0092&0.6361$\pm$0.0587&\textcolor{green!70!black}{\textbf{0.9837$\pm$0.0077}} &0.7030$\pm$0.0569&59.39$\pm$31.50
    \\
                       & CS$^{2}$-Net                     & 8.93       &              0.7747$\pm$0.0436& 0.9664$\pm$0.0120 &0.7524$\pm$0.0878& \textcolor{green!70!black}{\textbf{0.9855$\pm$0.0078}} &0.6344$\pm$0.0586& 0.9833$\pm$0.0083 &0.6941$\pm$0.0588&64.91$\pm$45.90
         \\
                       & MAA-Net                  & 5.80  & 0.7693$\pm$0.0508&0.9661$\pm$0.0120&0.7423$\pm$0.0969&\textcolor{green!70!black}{\textbf{0.9855$\pm$0.0091}} &0.6279$\pm$0.0669&0.9818$\pm$0.0095&0.69$\pm$0.0690&68.20$\pm$42.60
      \\
                       & FR-UNet       & 7.37 & \textcolor{green!70!black}{\textbf{0.7779$\pm$0.0459}}& \textcolor{green!70!black}{\textbf{0.9665$\pm$0.0118}} & 0.7646$\pm$0.0808 & 0.9841$\pm$0.0083& \textcolor{green!70!black}{\textbf{0.6376$\pm$0.0615}} & 0.9827$\pm$0.0078 & \textcolor{green!70!black}{\textbf{0.7112$\pm$0.0603}}& \textcolor{green!70!black}{\textbf{53.92$\pm$27.93}}
  \\ \hline
\multirow{6}{*}{3D}    & UNet    & 90.27                      & 0.7692$\pm$0.0490&0.9648$\pm$0.0121&0.7690$\pm$0.0887&0.9824$\pm$0.0090&0.6275$\pm$0.0650&\textcolor{blue}{\textbf{0.9818$\pm$0.0090}} &0.6975$\pm$0.0663& \textcolor{blue}{\textbf{62.95$\pm$37.21}}
    \\
                       & Res-UNet     & 26.38                      & \textcolor{blue}{\textbf{0.7701$\pm$0.0454}} &\textcolor{blue}{\textbf{0.9653$\pm$0.0121}} &0.7597$\pm$0.0809& \textcolor{blue}{\textbf{0.9835$\pm$0.0087}} & \textcolor{blue}{\textbf{0.6284$\pm$0.0606}} &0.9817$\pm$0.0092& \textcolor{blue}{\textbf{0.6991$\pm$0.0606}} &73.19$\pm$45.50       \\
                       & UNet++   & 26.17                      & 0.7649$\pm$0.0448&0.9639$\pm$0.0121& \textcolor{blue}{\textbf{0.7712$\pm$0.0791}} &0.9811$\pm$0.0088&0.6215$\pm$0.0589&0.9811$\pm$0.0092&0.6989$\pm$0.0595&67.03$\pm$48.51    \\
                       & Att-UNet    & 23.60   & 0.7677$\pm$0.0513& 0.9647$\pm$0.0131&0.7570$\pm$0.0816& 0.9829$\pm$0.0106& 0.6258$\pm$0.0686& 0.9811$\pm$0.0098& 0.6966$\pm$0.0624& 68.32$\pm$40.96 \\
                       & CS$^{2}$-Net   & 23.83  & 0.7687$\pm$0.0472&0.9645$\pm$0.0124& \textcolor{blue}{\textbf{0.7712$\pm$0.0838}} &0.9816$\pm$0.0099&0.6267$\pm$0.0628&0.9808$\pm$0.0100&0.6960$\pm$0.0650&67.91$\pm$45.25         \\
                       & FR-UNet & 20.57   & 0.7635$\pm$0.0587&0.9646$\pm$0.0123&0.7518$\pm$0.0889&0.9830$\pm$0.0107&0.6211$\pm$0.0754&0.9815$\pm$0.0096&0.6913$\pm$0.0686&65.65$\pm$45.09     \\ \hline
\multirow{5}{*}{3D$\rightarrow$2D} & IPN       & 6.36                       &0.7588$\pm$0.0506&0.9643$\pm$0.0119&0.7395$\pm$0.0908&0.9837$\pm$0.0100&0.6140$\pm$0.0651&0.9715$\pm$0.0200&0.6874$\pm$0.0740&61.85$\pm$36.92  \\
                       & PSC  & 6.43                       & 0.7649$\pm$0.0523&0.9646$\pm$0.0123&0.7552$\pm$0.0896&0.9832$\pm$0.0095&0.6223$\pm$0.069&0.9814$\pm$0.0104&0.6965$\pm$0.0604&76.05$\pm$44.63
   \\
                       & SVS-Net   & 8.48   &0.7697$\pm$0.0435&0.9659$\pm$0.0115&0.7474$\pm$0.0826&0.9853$\pm$0.0072&0.6277$\pm$0.0578&0.9803$\pm$0.0085&0.6880$\pm$0.0580&97.33$\pm$49.10   \\
                       & ST-UNet     & 104.37 & 0.7818$\pm$0.0406& \textcolor{red}{\textbf{0.9675$\pm$0.0119}} &0.7533$\pm$0.0821&\textcolor{red}{\textbf{0.9864$\pm$0.0070}} &0.6435$\pm$0.0551&0.9843$\pm$0.0075&0.7073$\pm$0.0587&62.54$\pm$38.25
      \\
                       & VSS-Net    & 18.69  & \textcolor{red}{\textbf{0.7822$\pm$0.0419}} & 0.9668 $\pm$ 0.0119 & \textcolor{red}{\textbf{0.7770$\pm$0.0830}} & 0.9838$\pm$0.0076          & \textcolor{red}{\textbf{0.6442$\pm$0.0568}} & \textcolor{red}{\textbf{0.9857$\pm$0.0070}} & \textcolor{red}{\textbf{0.7119$\pm$0.0583}} & \textcolor{red}{\textbf{43.02$\pm$27.47}} \\ \hline
\end{tabular}
\end{table*}

Table~\ref{table:model} demonstrates that the comprehensive performance of VSS-Net surpasses other networks, achieving the best DSC, Sen, IOU, AUC, clDice and VC, among all the evaluated methods. Using 2D FR-UNet as a baseline, VSS-Net shows improvements in numerous metrics except for Spe. Specifically, Sen has increased by 1.24\%. ST U-Net, also employing ConvGRU, shows performance very close to VSS-Net, achieving the best results in Acc and Spe. However, the number of model parameters is approximately 5.58 times that of VSS-Net. The VC score that is 19.52 higher than VSS-Net indicates poorer vascular connectivity. Both IPN and SVS-Net employ 3D convolutions in the encoder and 2D convolutions in the decoder, incorporating dimensionality reduction in the intermediate stages. PSC's segmentation results are similar to those of 3D models, but IPN shows poorer performance, possibly due to the absence of skip connections, which are significantly important for enhancing the detailing in vascular structures. 

\begin{figure*}[t]
\centering
\includegraphics[scale=0.53]{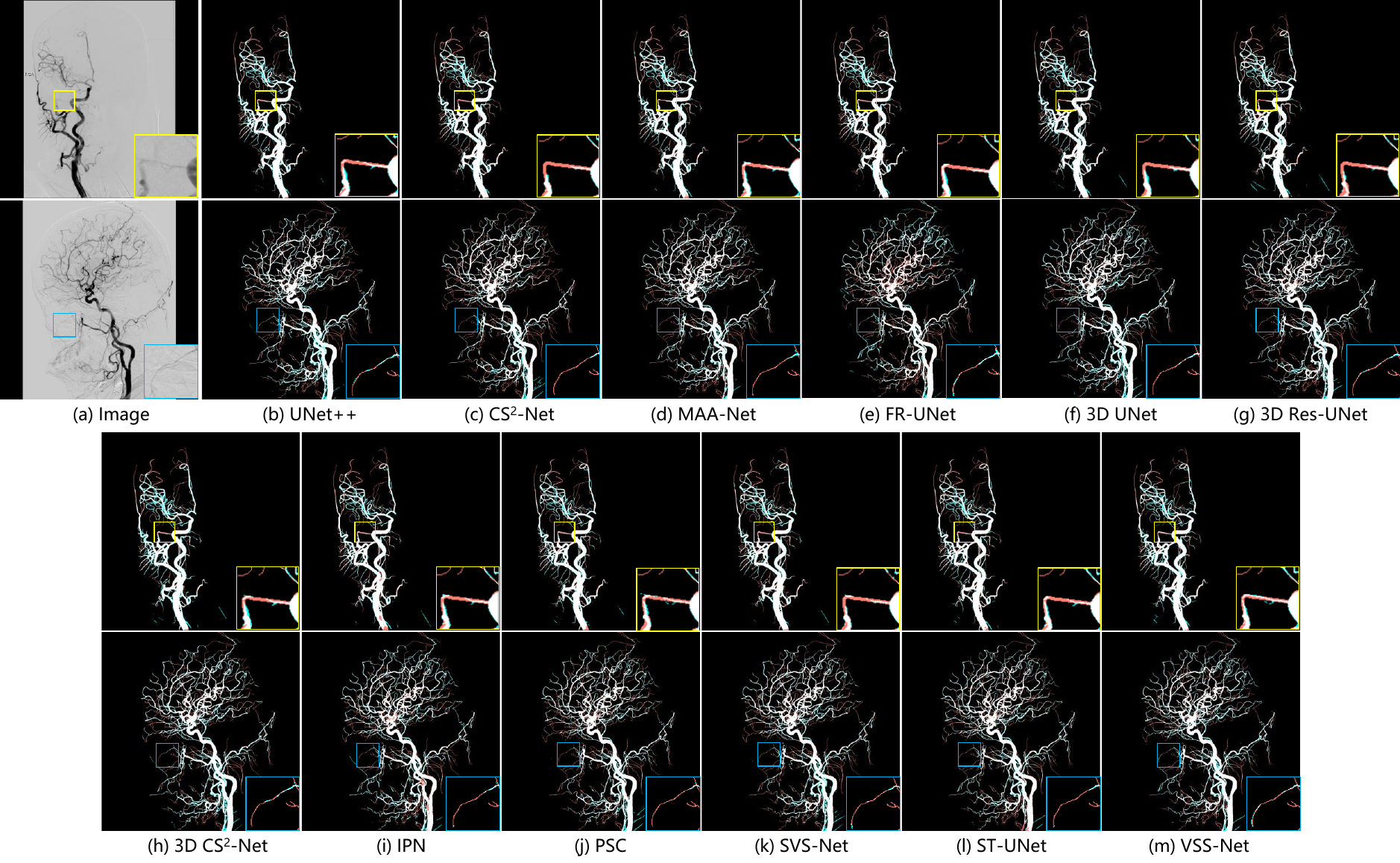}
\caption{Visualization of segmentation results of state-of-the-art models. The first and third rows display images of the anteroposterior view, while the second and fourth rows present images of the lateral view. In the segmentation maps for various methods, red pixels indicate false negatives, and green pixels signify false positives. The white and black pixels represent correctly segmented vessels and background, respectively. } \label{figure:vs}
\end{figure*}

Upon overall comparison of the DSC scores for all `DR+2D model' and `DR+3D model', 2D models slightly outperform 3D models with fewer parameters. Although in terms of data format, 2D+Time DSA sequences are equivalent to 3D data, they do not have a one-to-one spatial relationship with the ground truth. More importantly, 3D convolutions can only model local temporal representations. The `DR+2D model' benefits from treating `sequences' as `channels', which enables it to learn global representations along the sequence dimension while performing 2D convolutions, which aggregates contrast agent responses that reflect vascular morphological information in the sequence direction. However, the approach of setting fixed channel numbers is not applicable to variable image sequences. We have visualized the segmentation results of the methods that exhibited outstanding performance, as showed in Fig.~\ref{figure:vs}. In the figure, the red and blue colors represent pixels that are false negatives and false positives, respectively, indicative of segmentation errors. It can be observed that a prevalent issue shared by all methods is the abundant incorrect segmentation of thin vessels. Specifically, we enlarged representative details in the images, as illustrated by the yellow and blue boxes in Fig.~\ref{figure:vs}. In comparison with other methods, VSS-Net has the highest number of white pixels in the fine vessels, indicating the best segmentation results.

\begin{table*}[ht]
\centering
\caption{Results of Scribble-supervised IA sequence segmentation.} \label{table:wsl}
\renewcommand\arraystretch{1.5}
\tabcolsep5pt
\scriptsize
\begin{tabular}{lcccccccccccccccc}
\hline
\multicolumn{1}{c}{\multirow{2}{*}{Methods}} & \multicolumn{7}{c}{RDFA}                                                                                                                         & \multicolumn{7}{c}{SALE}                                                                                                  \\ \cline{2-17} 
\multicolumn{1}{c}{}                       & DSC$\uparrow$      & Acc$\uparrow$       & Sen$\uparrow$     & Spe$\uparrow$        & IOU$\uparrow$     & AUC$\uparrow$      & clDice$\uparrow$        & \multicolumn{1}{c|}{VC$\downarrow$}             & DSC$\uparrow$      & Acc$\uparrow$       & Sen$\uparrow$     & Spe$\uparrow$        & IOU$\uparrow$     & AUC$\uparrow$ & clDice$\uparrow$     & VC$\downarrow$            \\ \hline
pCE                                      & 0.7455          & 0.9586          & 0.7845          & 0.9735          & 0.5966          & 0.9379   &  0.7022     & \multicolumn{1}{c|}{60.11}          & 0.6396          & 0.9246          & 0.8462          & 0.9308          & 0.4738          & 0.9560 & 0.6589    & 10.02         \\
EM                                         & 0.7458          & 0.9575          & 0.8029          & 0.9707          & 0.5970          & 0.9725    &\textbf{0.7103}      & \multicolumn{1}{c|}{41.18}          & 0.6356          & 0.9226          & 0.8508          & 0.9281          & 0.4699          & 0.9618  & 0.6576       & 9.05          \\
RLoss                                       & 0.7146          & 0.9600          & 0.6539          & \textbf{0.9867} & 0.5581          & 0.9367  & 0.5906       & \multicolumn{1}{c|}{46.30}          & 0.5916          & 0.9147          & 0.7812          & 0.9257          & 0.4232          & 0.9475   & 0.5702       & \textbf{5.17} \\
    IVM                                       & 0.6606          & 0.9299          & \textbf{0.8546} & 0.9360          & 0.4971          & 0.9455          & 0.6910& \multicolumn{1}{c|}{71.58}          & 0.6322          & 0.9198          & \textbf{0.8760} & 0.9232          & 0.4662          & 0.9628    & \textbf{0.6738}      & 14.78         \\
DMPLS                                      & 0.7471          & 0.9581          & 0.7963          & 0.9717          & 0.5995          & 0.9708   &0.6996       & \multicolumn{1}{c|}{37.56}          & 0.6327          & 0.9208          & 0.8508          & 0.9259          & 0.4668          & 0.9643     &0.6601     & 9.86          \\
SSCR                                         & \textbf{0.7651} & \textbf{0.9639} & 0.7620          & 0.9810         & \textbf{0.6214} & \textbf{0.9778}   & 0.6977 & \multicolumn{1}{c|}{\textbf{34.48}} & \textbf{0.6574} & \textbf{0.9277} & 0.8398         & \textbf{0.9344} & \textbf{0.4958} & \textbf{0.9672} &  0.6732  &   6.47          \\ \hline
\end{tabular}
\end{table*}
\begin{figure*}[ht]
\centering
\includegraphics[scale=0.52]{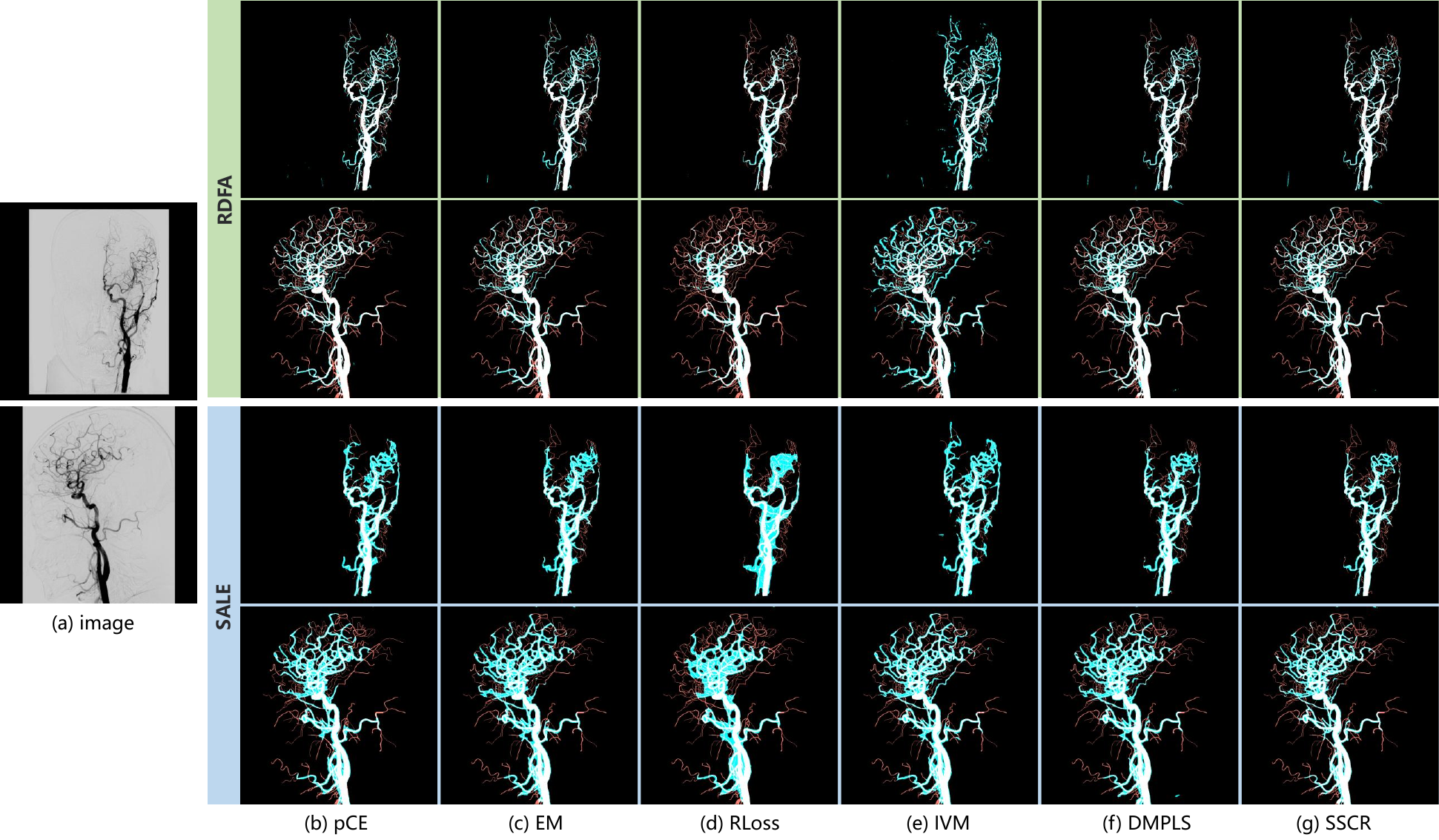}
\caption{Visualization of the segmentation results of scribble supervised segmentation methods trained on RDFA and SALE annotations.} \label{image:wsl}
\end{figure*}

\subsection{Weakly-supervised Segmentation}

We compared our method with five scribble-supervised segmentation methods: pCE~\cite{pce}, Entropy Minimization (EM)~\cite{em}, Regularized Loss (RLoss)~\cite{rloss}, Intra-class Intensity Variance Minimization (IVM)~\cite{ivm} and Dynamically Mixed Pseudo Labels Supervision (DMPLS)~\cite{luo2022scribble}. To ensure a fair comparison, all experimental networks employed the UNet~\cite{unet}, consistent with the approach in DMPLS. Table~\ref{table:wsl} presents the results of scribble-supervised segmentation of the DIAS dataset. Fig.~\ref{image:wsl} showcases the visual results of all methods. Thanks to pseudo supervision and consistency regularization for scribble supervision, SSCR achieved significantly better performance in RDFA and SALE by obtaining the highest scores in most metrics among all methods. It is noteworthy that in the segmentation results of IVM, the Sen on RDFA is 9.26\% higher compared to SSCR, and it is 3.62\% higher on SALE. Sen is used to measure the proportion of correctly identified positive classes (i.e., vessels) out of all actual positive classes, independent of the background. The visualization results from IVM indicate that achieving high Sen comes at the cost of introducing numerous incorrectly segmented vessel pixels. Consequently, this leads to lower scores in other evaluation metrics such as the DSC and AUC. Regarding SALE, the vessels segmented by the same method are thicker than the ones trained with RDFA annotation, leading to a larger number of false positive pixels at the vessel's edge. This outcome may have been caused by the SALE focusing on the annotation of thick vessels. Similarly, due to the lack of supervision for small vessels, where vascular connectivity issues are more prevalent, SALE struggles to extract fine vessels, resulting in a VC score significantly better than RDFA. From the last column of the figure, it is evident that SSCR's segmentation results have the least number of blue pixels, corroborating the best quantitative results presented in the Table~\ref{table:wsl}.

\begin{table}[ht]
\centering
\caption{Ablation studies results of SSCR for IA sequence segmentatio with SALE annotation. (pCE: partial Cross Entropy, PS: Pseudo-labeling Supervision, CPS: Cross Pseudo Supervision, CR: Consistency Regularization)} \label{table:wsl_as}
\renewcommand\arraystretch{1.5}
\tabcolsep2.4pt
\scriptsize

\begin{tabular}{cccccccccccc}
\hline
\multicolumn{4}{c}{Methods} & \multicolumn{7}{c}{Metrics}                                                                                               \\ \hline
pCE    & PS   & CPS   & CR   & DSC$\uparrow$      & Acc$\uparrow$       & Sen$\uparrow$     & Spe$\uparrow$        & IOU$\uparrow$     & AUC$\uparrow$       & clDice$\uparrow$   & VC$\downarrow$           \\ \hline
\checkmark       &      &       &      & 0.6396          & 0.9246          & 0.8462 & 0.9308          & 0.4738          & 0.9560     & 0.6589     & 10.02         \\
\checkmark       & \checkmark    &       &      & 0.6364          & 0.9244          & 0.8346          & 0.9316          & 0.4706          & 0.9479    &0.6437      & 6.57 \\
\checkmark       & \checkmark     & \checkmark      &      & 0.6472          & 0.9275          & 0.8385          &  \textbf{0.9346}          & 0.4827          & 0.9445     &0.6692     & 7.29          \\

\checkmark       & \checkmark    &       & \checkmark    & 0.6484          & 0.9276          & \textbf{0.8536} & 0.9334          & 0.4827          & 0.9639    &0.6685      & 8.28          \\

\checkmark       & \checkmark    & \checkmark      & \checkmark     & \textbf{0.6574} & \textbf{0.9277} & 0.8398         & 0.9344 & \textbf{0.4958} & \textbf{0.9672} &  \textbf{0.6732}  &   \textbf{6.47}    \\ \hline
\end{tabular}
\end{table}

\begin{table}[ht]
\centering
\caption{Ablation research results of RPST for semi-supervised IA sequence segmentation. (L: Labeled data, U: Unlabeled data) } \label{table:ssl}
\renewcommand\arraystretch{1.5}
\tabcolsep2.1pt
\scriptsize

\begin{tabular}{lcccccccccc}
\hline
\multicolumn{1}{c}{\multirow{2}{*}{Methods}} & \multicolumn{2}{c}{Data} & \multicolumn{7}{c}{Metrics}                                                                                                 \\ \cline{2-11} 
\multicolumn{1}{c}{}                         & L     & U    & DSC$\uparrow$      & Acc$\uparrow$       & Sen$\uparrow$     & Spe$\uparrow$        & IOU$\uparrow$     & AUC$\uparrow$    & clDice$\uparrow$           & VC$\downarrow$              \\ \hline
Teacher                                      & 1          & 0           & 0.6266          & 0.9259          & 0.6853          & 0.9451          & 0.4693          & 0.8538   &0.5173       & 385.62          \\
Student                                      & 1          & 30          & 0.6486          & 0.9294          & 0.7085          & 0.9464          & 0.4941          & 0.9505  &\textbf{0.5668}        & 107.35          \\
S w/o   SDA                            & 1          & 60          & 0.6394          & 0.9296          & 0.6922          & 0.9483          & 0.4832          & 0.9352   &0.5531       & 108.96          \\
Student                                      & 1          & 60          & \textbf{0.6502} & \textbf{0.9305} & \textbf{0.7177} & \textbf{0.9482} & \textbf{0.4958} & \textbf{0.9433} &0.5653& \textbf{103.09} \\ \hline
Teacher                                      & 3          & 0           & 0.6919          & 0.9490           & 0.7358          & 0.9673          & 0.5336          & 0.9182   &0.5921       & 320.02          \\
Student                                      & 3          & 30          & 0.7265          & 0.9555          & 0.7478          & 0.9721          & 0.5745          & 0.9591    & \textbf{0.6523}      & 121.84          \\
S w/o   SDA                            & 3          & 60          & 0.7228          & 0.9555          & 0.7456          & 0.9734          & 0.5700            & 0.9578   &0.6404       & 125.83          \\
Student                                      & 3          & 60          & \textbf{0.7275} & \textbf{0.9549} & \textbf{0.7595} & \textbf{0.9778} & \textbf{0.5893} & \textbf{0.9616} & 0.6502 & \textbf{125.78} \\ \hline
Teacher                                      & 10         & 0           & 0.7402          & 0.9596          & 0.7477          & 0.9779          & 0.5910           & 0.9694   &0.6512       & 148.89          \\
Student                                      & 10         & 30          & 0.7544          & 0.9618          & 0.7660           & 0.9786          & 0.6087          & 0.9769  &0.6747         & 93.31           \\
S w/o   SDA                            & 10         & 60          & 0.7556          & \textbf{0.9625} & 0.7552          & \textbf{0.9805} & 0.6096          & 0.9737   &0.6708       & 93.92           \\
Student                                      & 10         & 60          & \textbf{0.7579} & \textbf{0.9625} & \textbf{0.7662} & 0.9795          & \textbf{0.6130}  & \textbf{0.9786}& \textbf{0.6794} & \textbf{87.72}  \\ \hline
\end{tabular}
\end{table}

We conducted ablation studies to investigate the impact of various components in our proposed framework for weakly-supervised IA segmentation. As illustrated in Table~\ref{table:wsl_as}, with pCE serving as the baseline, we incrementally added components for comparative analysis. Initially, the introduction of pseudo-label supervision resulted in a slight decline in metrics such as Sen, AUC and clDice. Based on this, we utilized a dual-model framework to establish cross pseudo-supervision, leading to a 1.08\% enhancement in DSC. Compared to a single-model approach, a dual-model setup, receiving the same scribble supervision, generates pseudo-labels for mutual supervision. This avoids the scenario of a model being self-supervised by its own predictions, thereby enhancing the model's generalizability. We employ consistency regularization on the pseudo-labels generated by dual models to penalize inconsistent segmentation. Compared to `pCE+PS', `pCE+PS+CR' exhibits a 1.2\% improvement in DSC. Additionally, it has attained the highest Sen. We implement a combined strategy of cross pseudo-supervision and consistency regularization on the pseudo-labels produced by dual models. The integrated approach, termed as SSCR and denoted as `pCE+PS+CPS+CR', has achieved notable improvements, surpassing `pCE+PS+CPS' and `pCE+PS+CR' by 1.02\% and 0.9\%, respectively.


\subsection{Semi-supervised Segmentation}

The number of labeled and unlabeled samples and strong data augmentation (SDA) affect the performance of RPST. In this section, we randomly select  1, 3, and 10 labeled DSA sequences from the DIAS training set for SSS experiments. We leverage 30 and 60 DSA unlabeled sequences to generate pseudo labels and participate in training student. Table~\ref{table:ssl} shows that pseudo-labeled data generated by the trained teacher significantly improves the performance of the model, especially when there are fewer labeled data available. For example, the DSC of the student is improved by 2.2\% with one labeled data and 30 unlabeled data compared to the teacher. This performance gain is positively correlated with the amount of unlabeled data. Ablation studies on SDA demonstrate the effectiveness of enforcing consistency through the application of both strong and weak data augmentation. While the pseudo-labeled data used to train the student model contains noisy labels, they can still enhance the model's generalization capability, albeit with the supervision of incorrect pixel labels of vessel. The results of DIAS show that the benefits brought by the pseudo-labeled data outweigh the drawbacks of noisy labels in IA sequence segmentation.  As the number of labeled data increases, this beneficial effect gradually diminishes.

\begin{figure}[t]
\centering
\includegraphics[scale=0.39]{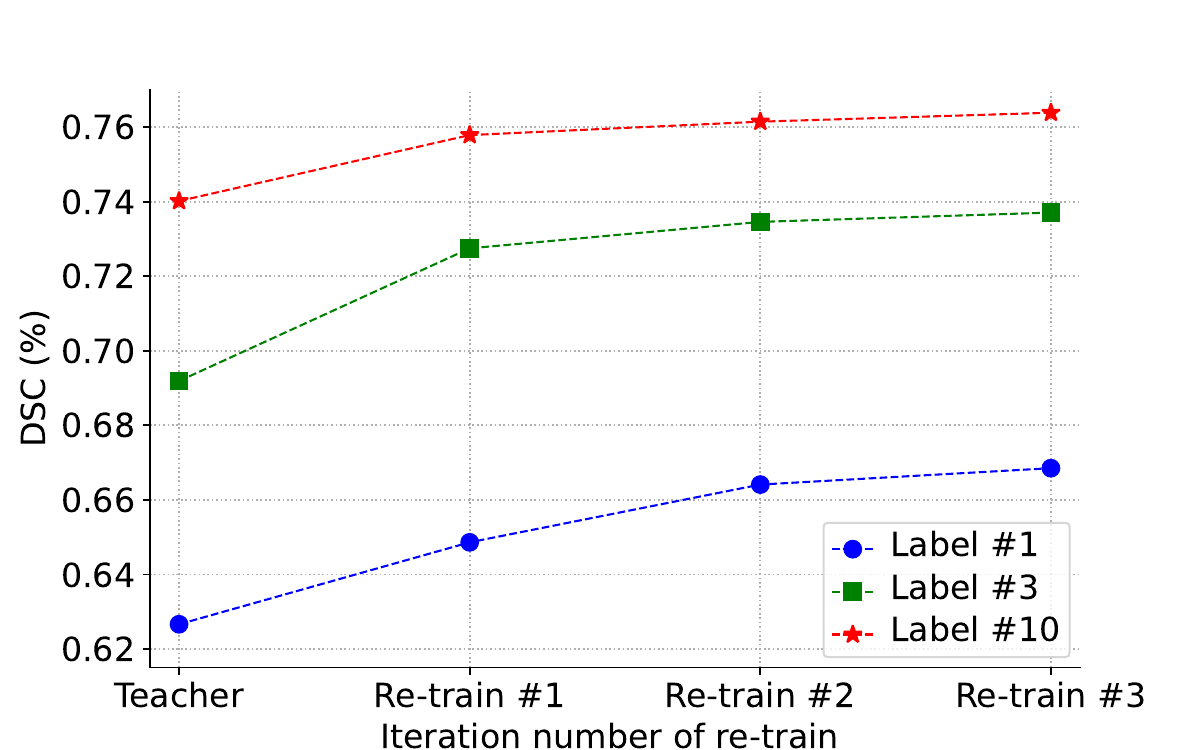}
\caption{Result of iterative training with different amounts of labeled data.} \label{image:ssl}
\end{figure}

All the unlabeled images can be labeled with a new teacher, i.e., the student from the previous stage, and re-trained in the iterative training stage. Fig.~\ref{image:ssl} presents our examination of the effectiveness of iterative training in RPST. In the extremely scarce-label regime with only one labeled sequence (\#1), performance is improved from 65.02\% to 66.85\% with the extra two-iteration training. The performance of \#10 improved from 75.79\% to 76.39\%. Each iteration enables the model to further learn and adjust based on the foundation established by previous iterations, increasing its confidence in its own predictions and resulting in the generation of more reliable pseudo-labels, thereby enhancing segmentation accuracy. With an increasing number of training iterations, the marginal improvement in performance diminishes.

\section{Discussion}

\subsection{Impact of Sequences as Inputs}
\label{sec:Impact of Sequences as Inputs}

\begin{figure*}[h]
\centering
\includegraphics[scale=0.6]{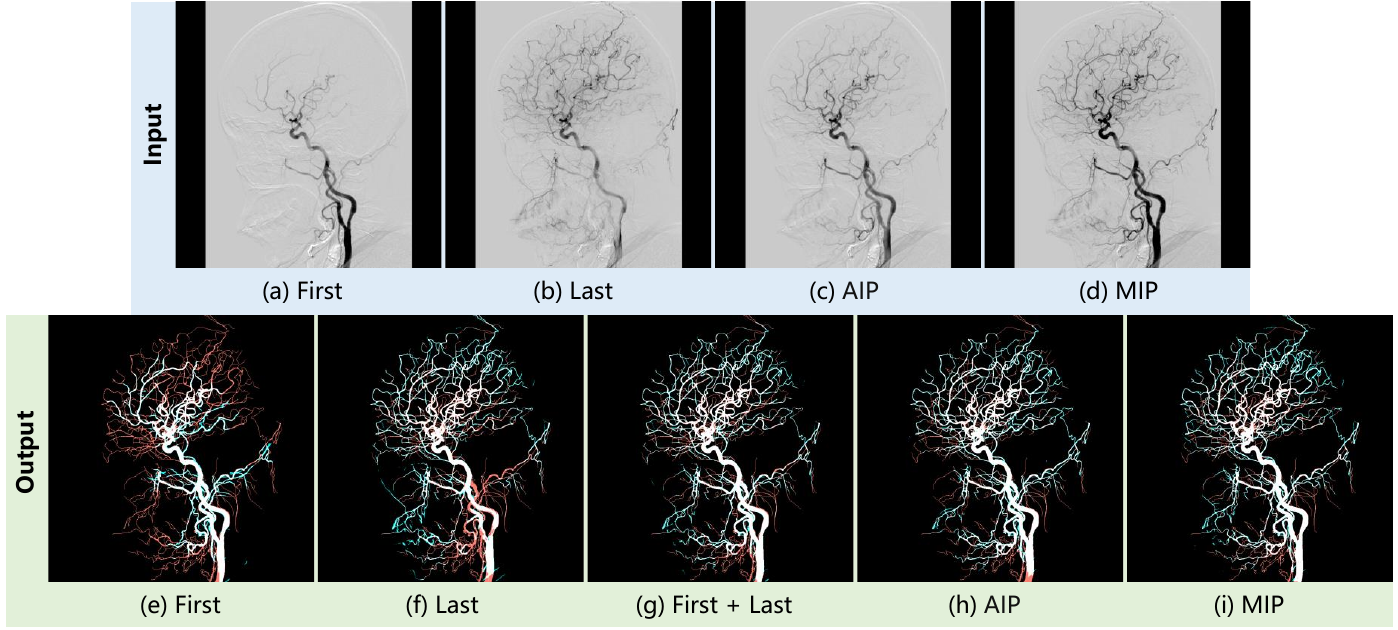}
\caption{The first row is the visualization of various inputs, and the second row is the corresponding segmentation result.} \label{figure:input}
\end{figure*}
The DSA sequence is a 2D temporal image sequence that represents the contrast agent's response at different times while flowing through the blood. Theoretically, accurate delineation of vessels requires the aggregation of contrast agent responses from all images in the sequence. In clinical practice, neurosurgeons commonly utilize Minimum Intensity Projection (MIP) to process target sequences in DSA for the visualization of vascular structures~\cite{MIP}. MIP is a straightforward and effective method for aggregating significant vascular information from target sequences. To evaluate the impact of various input modalities on segmentation performance, we conduct comparative experiments that include subsets of the image sequence (First, Last, and First+Last), MIP and Average Intensity Projection (AIP) image as input to the VSS-Net. As shown in Table~\ref{table:input}, In `First', where the contrast agent is exclusively displayed in the ICA, the segmentation results are inferior, the DSC is only 62.32\%. In comparison, the `Last' as input to the model obtained better results, thanks to the more angiographic information in thin vessel, the DSC is 76.04\%. The quantitative results are consistent with those portrayed in Fig.~\ref{figure:input} (e) and (f). When using only the first image as the model input, the large vessels are accurately segmented. However, due to the absence of angiographic information for fine vessels in the first image frame, the thin vessel at the arterial termini within the MCA remain incomplete. Conversely, using only the last image as input enables the accurate segmentation of most small vessels, yet it falls short in precisely delineating the larger vessels in proximity to the ICA. 

\begin{table}[t]
\centering
\caption{Results of fully-supervised segmentation with various input} \label{table:input}
\renewcommand\arraystretch{1.5}
\tabcolsep3.8pt
\scriptsize
\begin{tabular}{lcccccccc}
\hline
\multicolumn{1}{c}{\multirow{2}{*}{Input}} & \multicolumn{7}{c}{Metrics}                                                                                                \\ \cline{2-9} 
\multicolumn{1}{c}{}                       & DSC$\uparrow$      & Acc$\uparrow$       & Sen$\uparrow$     & Spe$\uparrow$        & IOU$\uparrow$     & AUC$\uparrow$     & clDice$\uparrow$  & VC$\downarrow$             \\ \hline
First    & 0.6232          & 0.9515          & 0.5388          & \textbf{0.9864} & 0.4601          & 0.9049   &0.6835        & \textbf{12.05} \\
Last & 0.7604          & 0.9647          & 0.7394          & 0.9842          & 0.6146 & 0.9813     &0.6940     & 51.89          \\
First+Last   & 0.7777          & 0.9668          & 0.7589          & 0.9850 & 0.6373          & 0.9846 &0.7039         & 44.22 \\
Mean                                       & 0.7790          & 0.9663          & 0.7695          & 0.9834          & 0.6398          & 0.9809   &0.6997       & 39.25          \\
Min                                        & 0.7802          & \textbf{0.9673} & 0.7631          & 0.9854          & 0.6418 & 0.9842       &0.7040   & 40.02          \\
Full                                       & \textbf{0.7822} & 0.9668          & \textbf{0.7770} & 0.9838          & \textbf{0.6442} & \textbf{0.9857}   &\textbf{0.7119} & 43.02          \\ \hline
\end{tabular}
\end{table}

The combination of early and late arterial angiography, known as `First+Last', yielded favorable results, surpassing those of both inputs used independently. As shown in Fig.~\ref{figure:input} (a) and (b), although the combination of the `First' and `Last' images from the arterial phase can display the majority of the vascular structures filled with contrast agent, we can still utilize the intermediate images to supplement details. The experimental findings indicate that employing the complete DSA sequence as input is imperative for achieving optimal performance. The segmentation results of MIP and AIP are quite close to those obtained using the method that utilizes the full DSA sequence as input. Density projection methods such as MIP and AIP convert arterial phase sequences into 2D images along the temporal dimension, which necessitates the pre-selection of target sequences to avoid the influence of capillary and venous phase angiography. Using DSA sequences as the network's input holds potential in addressing this challenge, as VSS-Net can selectively extract target sequence angiographic information from the entire series, including the capillary and venous phases, for the segmentation of IA.

\subsection{Potential Clinical Applications and Limitations}

Segmentation of vessels in DSA for quantifying their morphology has already found application in clinical research, particularly for planning interventional surgeries~\cite{spiegel20112d} and diagnosing strokes~\cite{tici}. However, as discussed in Section~\ref{sec:Impact of Sequences as Inputs}, the segmentation of a single DSA frame is inadequate for fully delineating the vascular structure. We believe that the application of VSS-Net for segmenting DSA vessel sequences in the aforementioned higher-level tasks will yield more accurate vascular structures. This advancement is expected to drive forward the progress of these higher-level tasks. Although the SSCR using scribble annotations falls short compared to fully supervised segmentation, the results shown in Fig.~\ref{image:wsl} reveal that this shortfall is primarily in the segmentation of smaller, terminal branches of vessels. Conversely, the segmentation of main vessels such as ICA, ACA, and MCA is comparatively effective. This is of clinical importance, as stroke-related ailments such as aneurysms~\cite{fully}, ICAS, and MCAO~\cite{occlusions} predominantly occur in these main vessels. It is noteworthy that our assessment indicates the time taken for scribble annotation of a DSA sequence is just one-tenth of that required for full annotation. In this work, we introduce the FSS, WSS, and SSS methods for intracranial artery segmentation in DSA sequences, specifically designed for the DIAS dataset. Importantly, the proposed approach is not confined to intracranial arteries. With the availability of relevant annotated data, it can be extended to other segmentation tasks in DSA sequences, such as venous segmentation~\cite{su2022spatio}, as well as for the diagnosis of stroke-related conditions like aneurysms, ICAS, and MCAO. 

The experimental results of the RPST demonstrates the potential of semi-supervised segmentation methods in the segmentation of IA in DSA sequences. One straightforward application is in assisting with data annotation. Currently, the DIAS dataset is relatively small, primarily due to the substantial effort expended on one-to-one pixel annotations of DSA sequences. In future work, we plan to leverage RPST to generate pseudo-labels for DSA sequences pending annotation. Annotation experts will then meticulously revise these pseudo-labels to obtain precise annotations, rather than starting the annotation process from the beginning. This strategy aims to increase annotation efficiency, thereby facilitating the further expansion of DIAS. On the other hand, in the DIAS dataset, intracranial arterial sequences are selected by neurosurgeons from DSA sequences that may include stages such as the capillary or venous phase.  Given the nature of 3D spatial projection, angiography from other phases in DSA may impact arterial segmentation, as the same pixel can represent an artery or a vein at different times. In our next phase of work, we plan to segment intracranial arteries from DSA sequences that include the venous phase and other stages, without relying on the neurosurgeons' prior knowledge to pre-select arterial phase sequences.

\section{Conclusion}

In this paper, we present the DSA-sequence IA segmentation dataset, DIAS, which provides pixel-level annotations along with two forms of weakly supervised annotations. We establish a benchmark for evaluating IA sequence segmentation on DIAS dataset, comprising (1) the vessel sequence segmentation network, adept at capturing the spatiotemporal representations in DSA angiography; (2) the weakly-supervised segmentation framework integrating scribble-supervision with consistency regularization; and (3) the random patch-based self-training framework, strategically designed to utilize unannotated DSA data to enhance segmentation performance. Leveraging the DIAS dataset, we have conducted comprehensive experiments to validate the effectiveness of our proposed methodologies, facilitating a deeper understanding of their performance characteristics and offering valuable insights for future research endeavors. Looking ahead, our research will continue, including the expansion of segmentation targets to veins and the segmentation of arteries from complete angiographic sequences. We would like to emphasize that the DIAS dataset and its associated benchmark will be made publicly available, enabling researchers to replicate our work and explore novel algorithms in this domain.

\appendices

\bibliographystyle{IEEEtran}
\bibliography{refer}

\begin{thebibliography}{10}
\providecommand{\url}[1]{#1}
\csname url@samestyle\endcsname
\providecommand{\newblock}{\relax}
\providecommand{\bibinfo}[2]{#2}
\providecommand{\BIBentrySTDinterwordspacing}{\spaceskip=0pt\relax}
\providecommand{\BIBentryALTinterwordstretchfactor}{4}
\providecommand{\BIBentryALTinterwordspacing}{\spaceskip=\fontdimen2\font plus
\BIBentryALTinterwordstretchfactor\fontdimen3\font minus
  \fontdimen4\font\relax}
\providecommand{\BIBforeignlanguage}[2]{{%
\expandafter\ifx\csname l@#1\endcsname\relax
\typeout{** WARNING: IEEEtran.bst: No hyphenation pattern has been}%
\typeout{** loaded for the language `#1'. Using the pattern for}%
\typeout{** the default language instead.}%
\else
\language=\csname l@#1\endcsname
\fi
#2}}
\providecommand{\BIBdecl}{\relax}
\BIBdecl

\bibitem{2022global}
M.~Vaduganathan, G.~A. Mensah, J.~V. Turco, V.~Fuster, and G.~A. Roth, ``The
  global burden of cardiovascular diseases and risk: a compass for future
  health,'' pp. 2361--2371, 2022.

\bibitem{imaging}
C.~P. Hess, ``Imaging in cerebrovascular disease,'' \emph{Molecular, Genetic,
  and Cellular Advances in Cerebrovascular Diseases}, pp. 1--23, 2018.

\bibitem{fully}
H.~Jin, J.~Geng, Y.~Yin, M.~Hu, G.~Yang, S.~Xiang, X.~Zhai, Z.~Ji, X.~Fan,
  P.~Hu \emph{et~al.}, ``Fully automated intracranial aneurysm detection and
  segmentation from digital subtraction angiography series using an end-to-end
  spatiotemporal deep neural network,'' \emph{Journal of NeuroInterventional
  Surgery}, vol.~12, no.~10, pp. 1023--1027, 2020.

\bibitem{dsa2022}
S.~Shaban, B.~Huasen, A.~Haridas, M.~Killingsworth, J.~Worthington, P.~Jabbour,
  and S.~M.~M. Bhaskar, ``Digital subtraction angiography in cerebrovascular
  disease: current practice and perspectives on diagnosis, acute treatment and
  prognosis,'' \emph{Acta Neurologica Belgica}, vol. 122, no.~3, pp. 763--780,
  2022.

\bibitem{tici}
R.~Su, S.~A. Cornelissen, M.~Van Der~Sluijs, A.~C. Van~Es, W.~H. Van~Zwam,
  D.~W. Dippel, G.~Lycklama, P.~J. Van~Doormaal, W.~J. Niessen, A.~Van Der~Lugt
  \emph{et~al.}, ``autotici: Automatic brain tissue reperfusion scoring on 2d
  dsa images of acute ischemic stroke patients,'' \emph{IEEE Transactions on
  Medical Imaging}, vol.~40, no.~9, pp. 2380--2391, 2021.

\bibitem{2d-3d-segmentation}
M.~Groher, F.~Bender, R.-T. Hoffmann, and N.~Navab, ``Segmentation-driven 2d-3d
  registration for abdominal catheter interventions,'' in \emph{International
  Conference on Medical Image Computing and Computer-Assisted
  Intervention}.\hskip 1em plus 0.5em minus 0.4em\relax Springer, 2007, pp.
  527--535.

\bibitem{spiegel20112d}
M.~Spiegel, T.~Redel, T.~Struffert, J.~Hornegger, and A.~Doerfler, ``A 2d
  driven 3d vessel segmentation algorithm for 3d digital subtraction
  angiography data,'' \emph{Physics in Medicine \& Biology}, vol.~56, no.~19,
  p. 6401, 2011.

\bibitem{su2022spatio}
R.~Su, M.~van~der Sluijs, S.~Cornelissen, P.~J. van Doormaal, R.~v.~d. Broek,
  W.~van Zwam, J.~Hofmeijer, D.~Ruijters, W.~Niessen, A.~van~der Lugt
  \emph{et~al.}, ``Spatio-temporal u-net for cerebral artery and vein
  segmentation in digital subtraction angiography,'' \emph{arXiv preprint
  arXiv:2208.02355}, 2022.

\bibitem{meng2020multiscale}
C.~Meng, K.~Sun, S.~Guan, Q.~Wang, R.~Zong, and L.~Liu, ``Multiscale dense
  convolutional neural network for dsa cerebrovascular segmentation,''
  \emph{Neurocomputing}, vol. 373, pp. 123--134, 2020.

\bibitem{li2020cau}
R.-Q. Li, G.-B. Bian, X.-H. Zhou, X.~Xie, Z.-L. Ni, and Z.~Hou, ``Cau-net: A
  novel convolutional neural network for coronary artery segmentation in
  digital substraction angiography,'' in \emph{International Conference on
  Neural Information Processing}.\hskip 1em plus 0.5em minus 0.4em\relax
  Springer, 2020, pp. 185--196.

\bibitem{weakly}
A.~Vepa, A.~Choi, N.~Nakhaei, W.~Lee, N.~Stier, A.~Vu, G.~Jenkins, X.~Yang,
  M.~Shergill, M.~Desphy \emph{et~al.}, ``Weakly-supervised convolutional
  neural networks for vessel segmentation in cerebral angiography,'' in
  \emph{Proceedings of the IEEE/CVF Winter Conference on Applications of
  Computer Vision}, 2022, pp. 585--594.

\bibitem{phtrans}
W.~Liu, T.~Tian, W.~Xu, H.~Yang, X.~Pan, S.~Yan, and L.~Wang, ``Phtrans:
  Parallelly aggregating global and local representations for medical image
  segmentation,'' in \emph{Medical Image Computing and Computer Assisted
  Intervention--MICCAI 2022: 25th International Conference, Singapore,
  September 18--22, 2022, Proceedings, Part V}.\hskip 1em plus 0.5em minus
  0.4em\relax Springer, 2022, pp. 235--244.

\bibitem{fr-unet}
W.~Liu, H.~Yang, T.~Tian, Z.~Cao, X.~Pan, W.~Xu, Y.~Jin, and F.~Gao,
  ``Full-resolution network and dual-threshold iteration for retinal vessel and
  coronary angiograph segmentation,'' \emph{IEEE Journal of Biomedical and
  Health Informatics}, vol.~26, no.~9, pp. 4623--4634, 2022.

\bibitem{maa-unet}
W.~Liu, H.~Yang, T.~Tian, X.~Pan, and W.~Xu, ``Multiscale attention aggregation
  network for 2d vessel segmentation,'' in \emph{ICASSP 2022-2022 IEEE
  International Conference on Acoustics, Speech and Signal Processing
  (ICASSP)}.\hskip 1em plus 0.5em minus 0.4em\relax IEEE, 2022, pp. 1436--1440.

\bibitem{DRIVE}
\BIBentryALTinterwordspacing
J.~Staal, M.~D. Abr{\`{a}}moff, M.~Niemeijer, M.~A. Viergever, and B.~van
  Ginneken, ``Ridge-based vessel segmentation in color images of the retina,''
  \emph{{IEEE} Trans. Medical Imaging}, vol.~23, no.~4, pp. 501--509, 2004.
  [Online]. Available: \url{https://doi.org/10.1109/TMI.2004.825627}
\BIBentrySTDinterwordspacing

\bibitem{CHASEDB}
M.~M. {Fraz}, P.~{Remagnino}, A.~{Hoppe}, B.~{Uyyanonvara}, A.~R. {Rudnicka},
  C.~G. {Owen}, and S.~A. {Barman}, ``An ensemble classification-based approach
  applied to retinal blood vessel segmentation,'' \emph{IEEE Transactions on
  Biomedical Engineering}, vol.~59, no.~9, pp. 2538--2548, Sep. 2012.

\bibitem{STARE}
A.~Hoover, V.~Kouznetsova, and M.~Goldbaum, ``Locating blood vessels in retinal
  images by piecewise threshold probing of a matched filter response,''
  \emph{IEEE Transactions on Medical imaging}, vol.~19, no.~3, pp. 203--210,
  2000.

\bibitem{HRF}
J.~Odstrcilik, R.~Kolar, A.~Budai, J.~Hornegger, J.~Jan, J.~Gazarek, T.~Kubena,
  P.~Cernosek, O.~Svoboda, and E.~Angelopoulou, ``Retinal vessel segmentation
  by improved matched filtering: evaluation on a new high-resolution fundus
  image database,'' \emph{IET Image Processing}, vol.~7, no.~4, pp. 373--383,
  2013.

\bibitem{ROSE}
Y.~Ma, H.~Hao, J.~Xie, H.~Fu, J.~Zhang, J.~Yang, Z.~Wang, J.~Liu, Y.~Zheng, and
  Y.~Zhao, ``Rose: a retinal oct-angiography vessel segmentation dataset and
  new model,'' \emph{IEEE transactions on medical imaging}, vol.~40, no.~3, pp.
  928--939, 2020.

\bibitem{CHUAC}
\BIBentryALTinterwordspacing
A.~Carballal, F.~J. N{\'{o}}voa, C.~Fernandez{-}Lozano,
  M.~Garc{\'{\i}}a{-}Guimaraes, G.~L{\'{o}}pez{-}Campos,
  R.~Calvi{\~{n}}o{-}Santos, J.~M. V{\'{a}}zquez{-}Naya, and A.~Pazos,
  ``Automatic multiscale vascular image segmentation algorithm for coronary
  angiography,'' \emph{Biomed. Signal Process. Control.}, vol.~46, pp. 1--9,
  2018. [Online]. Available: \url{https://doi.org/10.1016/j.bspc.2018.06.007}
\BIBentrySTDinterwordspacing

\bibitem{DCA1}
\BIBentryALTinterwordspacing
F.~Cervantes-Sanchez, I.~Cruz-Aceves, A.~Hernandez-Aguirre, M.~A.
  Hernandez-Gonzalez, and S.~E. Solorio-Meza, ``{Automatic Segmentation of
  Coronary Arteries in X-ray Angiograms using Multiscale Analysis and
  Artificial Neural Networks},'' \emph{Applied Sciences}, vol.~9, no.~24, 2019.
  [Online]. Available: \url{https://www.mdpi.com/2076-3417/9/24/5507}
\BIBentrySTDinterwordspacing

\bibitem{ASOCA}
R.~Gharleghi, D.~Adikari, K.~Ellenberger, S.-Y. Ooi, C.~Ellis, C.-M. Chen,
  R.~Gao, Y.~He, R.~Hussain, C.-Y. Lee \emph{et~al.}, ``Automated segmentation
  of normal and diseased coronary arteries--the asoca challenge,''
  \emph{Computerized Medical Imaging and Graphics}, vol.~97, p. 102049, 2022.

\bibitem{VESSEL12}
R.~D. Rudyanto, S.~Kerkstra, E.~M. Van~Rikxoort, C.~Fetita, P.-Y. Brillet,
  C.~Lefevre, W.~Xue, X.~Zhu, J.~Liang, I.~{\"O}ks{\"u}z \emph{et~al.},
  ``Comparing algorithms for automated vessel segmentation in computed
  tomography scans of the lung: the vessel12 study,'' \emph{Medical image
  analysis}, vol.~18, no.~7, pp. 1217--1232, 2014.

\bibitem{BEFD}
M.~Zhang, F.~Yu, J.~Zhao, L.~Zhang, and Q.~Li, ``Befd: boundary enhancement and
  feature denoising for vessel segmentation,'' in \emph{International
  Conference on Medical Image Computing and Computer-Assisted
  Intervention}.\hskip 1em plus 0.5em minus 0.4em\relax Springer, 2020, pp.
  775--785.

\bibitem{zhang2018vesselness}
J.~Zhang, G.~Wang, H.~Xie, S.~Zhang, Z.~Shi, and L.~Gu,
  ``Vesselness-constrained robust pca for vessel enhancement in x-ray coronary
  angiograms,'' \emph{Physics in Medicine \& Biology}, vol.~63, no.~15, p.
  155019, 2018.

\bibitem{sang2007knowledge}
N.~Sang, H.~Li, W.~Peng, and T.~Zhang, ``Knowledge-based adaptive thresholding
  segmentation of digital subtraction angiography images,'' \emph{Image and
  Vision Computing}, vol.~25, no.~8, pp. 1263--1270, 2007.

\bibitem{liu2018vessel}
B.~Liu, Q.~Jiang, W.~Liu, M.~Wang, S.~Zhang, X.~Zhang, B.~Zhang, and Z.~Yue,
  ``A vessel segmentation method for serialized cerebralvascular dsa images
  based on spatial feature point set of rotating coordinate system,''
  \emph{Computer Methods and Programs in Biomedicine}, vol. 161, pp. 55--72,
  2018.

\bibitem{unet}
O.~Ronneberger, P.~Fischer, and T.~Brox, ``U-net: Convolutional networks for
  biomedical image segmentation,'' in \emph{International Conference on Medical
  image computing and computer-assisted intervention}.\hskip 1em plus 0.5em
  minus 0.4em\relax Springer, 2015, pp. 234--241.

\bibitem{scsnet}
H.~Wu, W.~Wang, J.~Zhong, B.~Lei, Z.~Wen, and J.~Qin, ``Scs-net: A scale and
  context sensitive network for retinal vessel segmentation,'' \emph{Medical
  Image Analysis}, vol.~70, p. 102025, 2021.

\bibitem{wu2019vessel}
Y.~Wu, Y.~Xia, Y.~Song, D.~Zhang, D.~Liu, C.~Zhang, and W.~Cai, ``Vessel-net:
  Retinal vessel segmentation under multi-path supervision,'' in
  \emph{International conference on medical image computing and
  computer-assisted intervention}.\hskip 1em plus 0.5em minus 0.4em\relax
  Springer, 2019, pp. 264--272.

\bibitem{cs2-net}
L.~Mou, Y.~Zhao, H.~Fu, Y.~Liu, J.~Cheng, Y.~Zheng, P.~Su, J.~Yang, L.~Chen,
  A.~F. Frangi \emph{et~al.}, ``Cs2-net: Deep learning segmentation of
  curvilinear structures in medical imaging,'' \emph{Medical image analysis},
  vol.~67, p. 101874, 2021.

\bibitem{agnet}
S.~Zhang, H.~Fu, Y.~Yan, Y.~Zhang, Q.~Wu, M.~Yang, M.~Tan, and Y.~Xu,
  ``Attention guided network for retinal image segmentation,'' in
  \emph{International conference on medical image computing and
  computer-assisted intervention}.\hskip 1em plus 0.5em minus 0.4em\relax
  Springer, 2019, pp. 797--805.

\bibitem{rvsegnet}
W.~Wang, J.~Zhong, H.~Wu, Z.~Wen, and J.~Qin, ``Rvseg-net: An efficient feature
  pyramid cascade network for retinal vessel segmentation,'' in
  \emph{International Conference on Medical Image Computing and
  Computer-Assisted Intervention}.\hskip 1em plus 0.5em minus 0.4em\relax
  Springer, 2020, pp. 796--805.

\bibitem{rvgan}
S.~A. Kamran, K.~F. Hossain, A.~Tavakkoli, S.~L. Zuckerbrod, K.~M. Sanders, and
  S.~A. Baker, ``Rv-gan: segmenting retinal vascular structure in fundus
  photographs using a novel multi-scale generative adversarial network,'' in
  \emph{International Conference on Medical Image Computing and
  Computer-Assisted Intervention}.\hskip 1em plus 0.5em minus 0.4em\relax
  Springer, 2021, pp. 34--44.

\bibitem{gnn}
S.~Y. Shin, S.~Lee, I.~D. Yun, and K.~M. Lee, ``Deep vessel segmentation by
  learning graphical connectivity,'' \emph{Medical image analysis}, vol.~58, p.
  101556, 2019.

\bibitem{sgl}
Y.~Zhou, H.~Yu, and H.~Shi, ``Study group learning: Improving retinal vessel
  segmentation trained with noisy labels,'' in \emph{International Conference
  on Medical Image Computing and Computer-Assisted Intervention}.\hskip 1em
  plus 0.5em minus 0.4em\relax Springer, 2021, pp. 57--67.

\bibitem{boosting}
R.~Xu, T.~Liu, X.~Ye, L.~Lin, and Y.-W. Chen, ``Boosting connectivity in
  retinal vessel segmentation via a recursive semantics-guided network,'' in
  \emph{International Conference on Medical Image Computing and
  Computer-Assisted Intervention}.\hskip 1em plus 0.5em minus 0.4em\relax
  Springer, 2020, pp. 786--795.

\bibitem{loss}
H.~Zhao, H.~Li, and L.~Cheng, ``Improving retinal vessel segmentation with
  joint local loss by matting,'' \emph{Pattern Recognition}, vol.~98, p.
  107068, 2020.

\bibitem{svsnet}
D.~Hao, S.~Ding, L.~Qiu, Y.~Lv, B.~Fei, Y.~Zhu, and B.~Qin, ``Sequential vessel
  segmentation via deep channel attention network,'' \emph{Neural Networks},
  vol. 128, pp. 172--187, 2020.

\bibitem{chen2021seminar}
H.~Chen, J.~Wang, H.~C. Chen, X.~Zhen, F.~Zheng, R.~Ji, and L.~Shao, ``Seminar
  learning for click-level weakly supervised semantic segmentation,'' in
  \emph{Proceedings of the IEEE/CVF International Conference on Computer
  Vision}, 2021, pp. 6920--6929.

\bibitem{luo2022scribble}
X.~Luo, M.~Hu, W.~Liao, S.~Zhai, T.~Song, G.~Wang, and S.~Zhang,
  ``Scribble-supervised medical image segmentation via dual-branch network and
  dynamically mixed pseudo labels supervision,'' in \emph{International
  Conference on Medical Image Computing and Computer-Assisted
  Intervention}.\hskip 1em plus 0.5em minus 0.4em\relax Springer, 2022, pp.
  528--538.

\bibitem{st++}
L.~Yang, W.~Zhuo, L.~Qi, Y.~Shi, and Y.~Gao, ``St++: Make self-training work
  better for semi-supervised semantic segmentation,'' in \emph{Proceedings of
  the IEEE/CVF Conference on Computer Vision and Pattern Recognition}, 2022,
  pp. 4268--4277.

\bibitem{flare22}
W.~Liu, W.~Xu, S.~Yan, L.~Wang, H.~Li, and H.~Yang, ``Combining self-training
  and hybrid architecture for semi-supervised abdominal organ segmentation,''
  in \emph{Fast and Low-Resource Semi-supervised Abdominal Organ Segmentation:
  MICCAI 2022 Challenge, FLARE 2022, Held in Conjunction with MICCAI 2022,
  Singapore, September 22, 2022, Proceedings}.\hskip 1em plus 0.5em minus
  0.4em\relax Springer, 2023, pp. 281--292.

\bibitem{cps}
X.~Chen, Y.~Yuan, G.~Zeng, and J.~Wang, ``Semi-supervised semantic segmentation
  with cross pseudo supervision,'' in \emph{Proceedings of the IEEE/CVF
  Conference on Computer Vision and Pattern Recognition}, 2021, pp. 2613--2622.

\bibitem{ouali2020semi}
Y.~Ouali, C.~Hudelot, and M.~Tami, ``Semi-supervised semantic segmentation with
  cross-consistency training,'' in \emph{Proceedings of the IEEE/CVF Conference
  on Computer Vision and Pattern Recognition}, 2020, pp. 12\,674--12\,684.

\bibitem{gu2022contrastive}
R.~Gu, J.~Zhang, G.~Wang, W.~Lei, T.~Song, X.~Zhang, K.~Li, and S.~Zhang,
  ``Contrastive semi-supervised learning for domain adaptive segmentation
  across similar anatomical structures,'' \emph{IEEE Transactions on Medical
  Imaging}, vol.~42, no.~1, pp. 245--256, 2022.

\bibitem{a2v}
\BIBentryALTinterwordspacing
F.~Galati, D.~Falcetta, R.~Cortese, B.~Casolla, F.~Prados, N.~Burgos, and M.~A.
  Zuluaga, ``{A2V:} {A} semi-supervised domain adaptation framework for brain
  vessel segmentation via two-phase training angiography-to-venography
  translation,'' in \emph{34th British Machine Vision Conference 2023, {BMVC}
  2023, Aberdeen, UK, November 20-24, 2023}.\hskip 1em plus 0.5em minus
  0.4em\relax {BMVA} Press, 2023, pp. 750--751. [Online]. Available:
  \url{http://proceedings.bmvc2023.org/750/}
\BIBentrySTDinterwordspacing

\bibitem{stylegan2}
T.~Karras, S.~Laine, M.~Aittala, J.~Hellsten, J.~Lehtinen, and T.~Aila,
  ``Analyzing and improving the image quality of stylegan,'' in
  \emph{Proceedings of the IEEE/CVF conference on computer vision and pattern
  recognition}, 2020, pp. 8110--8119.

\bibitem{DBLP:conf/iccv/ButoiOMSGD23}
\BIBentryALTinterwordspacing
V.~I. Butoi, J.~J.~G. Ortiz, T.~Ma, M.~R. Sabuncu, J.~V. Guttag, and A.~V.
  Dalca, ``Universeg: Universal medical image segmentation,'' in
  \emph{{IEEE/CVF} International Conference on Computer Vision, {ICCV} 2023,
  Paris, France, October 1-6, 2023}.\hskip 1em plus 0.5em minus 0.4em\relax
  {IEEE}, 2023, pp. 21\,381--21\,394. [Online]. Available:
  \url{https://doi.org/10.1109/ICCV51070.2023.01960}
\BIBentrySTDinterwordspacing

\bibitem{zhang2023challenges}
S.~Zhang and D.~Metaxas, ``On the challenges and perspectives of foundation
  models for medical image analysis,'' \emph{Medical Image Analysis}, p.
  102996, 2023.

\bibitem{rloss}
M.~Tang, F.~Perazzi, A.~Djelouah, I.~Ben~Ayed, C.~Schroers, and Y.~Boykov, ``On
  regularized losses for weakly-supervised cnn segmentation,'' in
  \emph{Proceedings of the European Conference on Computer Vision (ECCV)},
  2018, pp. 507--522.

\bibitem{ivm}
X.~Luo, W.~Liao, J.~Xiao, J.~Chen, T.~Song, X.~Zhang, K.~Li, D.~N. Metaxas,
  G.~Wang, and S.~Zhang, ``Word: A large scale dataset, benchmark and clinical
  applicable study for abdominal organ segmentation from ct image,''
  \emph{Medical Image Analysis}, vol.~82, p. 102642, 2022.

\bibitem{PSC}
D.~Lachinov, P.~Seeb{\"o}ck, J.~Mai, F.~Goldbach, U.~Schmidt-Erfurth, and
  H.~Bogunovic, ``Projective skip-connections for segmentation along a subset
  of dimensions in retinal oct,'' in \emph{International Conference on Medical
  Image Computing and Computer-Assisted Intervention}.\hskip 1em plus 0.5em
  minus 0.4em\relax Springer, 2021, pp. 431--441.

\bibitem{convgru}
\BIBentryALTinterwordspacing
N.~Ballas, L.~Yao, C.~Pal, and A.~C. Courville, ``Delving deeper into
  convolutional networks for learning video representations,'' in \emph{4th
  International Conference on Learning Representations, {ICLR} 2016, San Juan,
  Puerto Rico, May 2-4, 2016, Conference Track Proceedings}, Y.~Bengio and
  Y.~LeCun, Eds., 2016. [Online]. Available:
  \url{http://arxiv.org/abs/1511.06432}
\BIBentrySTDinterwordspacing

\bibitem{DB-convgru}
P.~Yan, G.~Li, Y.~Xie, Z.~Li, C.~Wang, T.~Chen, and L.~Lin, ``Semi-supervised
  video salient object detection using pseudo-labels,'' in \emph{Proceedings of
  the IEEE/CVF international conference on computer vision}, 2019, pp.
  7284--7293.

\bibitem{pseudo}
D.-H. Lee \emph{et~al.}, ``Pseudo-label: The simple and efficient
  semi-supervised learning method for deep neural networks,'' in \emph{Workshop
  on challenges in representation learning, ICML}, vol.~3, no.~2, 2013, p. 896.

\bibitem{adamw}
I.~Loshchilov and F.~Hutter, ``Decoupled weight decay regularization,''
  \emph{arXiv preprint arXiv:1711.05101}, 2017.

\bibitem{cldice}
S.~Shit, J.~C. Paetzold, A.~Sekuboyina, I.~Ezhov, A.~Unger, A.~Zhylka, J.~P.
  Pluim, U.~Bauer, and B.~H. Menze, ``cldice-a novel topology-preserving loss
  function for tubular structure segmentation,'' in \emph{Proceedings of the
  IEEE/CVF conference on computer vision and pattern recognition}, 2021, pp.
  16\,560--16\,569.

\bibitem{att-unet}
J.~Schlemper, O.~Oktay, M.~Schaap, M.~Heinrich, B.~Kainz, B.~Glocker, and
  D.~Rueckert, ``Attention gated networks: Learning to leverage salient regions
  in medical images,'' \emph{Medical image analysis}, vol.~53, pp. 197--207,
  2019.

\bibitem{unet++}
Z.~Zhou, M.~M. Rahman~Siddiquee, N.~Tajbakhsh, and J.~Liang, ``Unet++: A nested
  u-net architecture for medical image segmentation,'' in \emph{Deep learning
  in medical image analysis and multimodal learning for clinical decision
  support}.\hskip 1em plus 0.5em minus 0.4em\relax Springer, 2018, pp. 3--11.

\bibitem{res-unet}
X.~Xiao, S.~Lian, Z.~Luo, and S.~Li, ``Weighted res-unet for high-quality
  retina vessel segmentation,'' in \emph{2018 9th international conference on
  information technology in medicine and education (ITME)}.\hskip 1em plus
  0.5em minus 0.4em\relax IEEE, 2018, pp. 327--331.

\bibitem{IPN}
M.~Li, Y.~Chen, Z.~Ji, K.~Xie, S.~Yuan, Q.~Chen, and S.~Li, ``Image projection
  network: 3d to 2d image segmentation in octa images,'' \emph{IEEE
  Transactions on Medical Imaging}, vol.~39, no.~11, pp. 3343--3354, 2020.

\bibitem{pce}
M.~Tang, A.~Djelouah, F.~Perazzi, Y.~Boykov, and C.~Schroers, ``Normalized cut
  loss for weakly-supervised cnn segmentation,'' in \emph{Proceedings of the
  IEEE conference on computer vision and pattern recognition}, 2018, pp.
  1818--1827.

\bibitem{em}
Y.~Grandvalet and Y.~Bengio, ``Semi-supervised learning by entropy
  minimization,'' \emph{Advances in neural information processing systems},
  vol.~17, 2004.

\bibitem{MIP}
Y.~Wu, F.~Li, Y.~Wang, T.~Hu, and L.~Xiao, ``Utility of minimum intensity
  projection images based on three-dimensional cube t 1 weighted imaging for
  evaluating middle cerebral artery stenosis,'' \emph{The British Journal of
  Radiology}, vol.~94, no. 1121, p. 20210145, 2021.

\bibitem{occlusions}
J.~Khankari, Y.~Yu, J.~Ouyang, R.~Hussein, H.~M. Do, J.~J. Heit, and
  G.~Zaharchuk, ``Automated detection of arterial landmarks and vascular
  occlusions in patients with acute stroke receiving digital subtraction
  angiography using deep learning,'' \emph{Journal of NeuroInterventional
  Surgery}, vol.~15, no.~6, pp. 521--525, 2023.

\end{thebibliography}

\end{document}